\documentclass[aip,author-year]{revtex4-1}

\usepackage{graphicx}
\usepackage{dcolumn}
\usepackage{bm}
\usepackage[mathlines]{lineno}
\usepackage[colorlinks=true,citecolor=black,linkcolor=black]{hyperref}
\usepackage[utf8]{inputenc}
\usepackage[T1]{fontenc}
\usepackage{mathptmx}
\usepackage{etoolbox}
\usepackage{rotating}

\makeatletter
\def\@email#1#2{%
 \endgroup
 \patchcmd{\titleblock@produce}
  {\frontmatter@RRAPformat}
  {\frontmatter@RRAPformat{\produce@RRAP{*#1\href{mailto:#2}{#2}}}\frontmatter@RRAPformat}
  {}{}
}%
\makeatother
\begin{document}

\preprint{Physics of Fluids}

\title[Hydrodynamics of heterogeneous coral reefs]{Should we care about the spatial heterogeneity in coral reefs under unidirectional turbulent flows?}
\author{Akshay Patil}
 \affiliation{3D Geoinformation Research Group, Faculty of Architecture and the Built Environment, Delft University of Technology, Delft, The Netherlands}
 \email{a.l.patil@tudelft.nl}
\author{Clara Garc\'{i}a-S\'{a}nchez}%
\affiliation{3D Geoinformation Research Group, Faculty of Architecture and the Built Environment, Delft University of Technology, Delft, The Netherlands}

\date{\today}


\begin{abstract}
In this work, we systematically investigate the similarities and differences observed between a hydraulically rough wall comprised of an array of cylinders, massive corals, and branching corals arranged in a staggered manner, along with a stochastically generated coral bed using a scale-resolving computational framework.
Our data suggests that for all the flow parameters of interest, there is a substantial difference observed between the stochastic coral bed and the regularly arranged coral bed.
By analysing the double-averaged statistics and time-averaged spatial heterogeneity in the hydrodynamic response, we explain the differences observed between the four cases that bring out significant local effects.
These observations have important consequences for modelling coral-like roughness in numerical and experimental settings to better understand the mean flow statistics and the spatial heterogeneity induced as a consequence of the underlying coral geometry.
Our results can help inform the coastal ocean modelling efforts to further improve the inclusion of coral heterogeneity within two-equation closure models by further investigating the impact of spatially stochastic, rough bottom boundary layers.

\end{abstract}

\maketitle

\section{Introduction}

Turbulent flows through complex and multi-scale roughness canopies are observed in mechanical, atmospheric, fluvial, and coastal environments \citep{Hamilton2024,Kelly2023,Nepf2012,Anderson2011,Ghisalberti2002}, thus motivating the need to characterise the hydrodynamics for such complex flows. 
One of the peculiar aspects of flow through complex roughness canopies is the roughness-induced heterogeneity that introduces a complex mean flow response, thus giving rise to the so-called dispersive stress \citep{Jelly2018,Poggi2008}. 
Suppose the velocity vector $U_i(x_i,t)$, is partitioned as follows,

\begin{equation} \label{eq:velocity_Breakdown}
    U_i(x_i,t) \equiv \overline{U}_i(x_i) + u^{\prime}_i (x_i,t) \equiv \langle \overline{U}_i\rangle (x_3) + \widetilde{U}_i(x_i) + u^{\prime}_i (x_i,t),
\end{equation}

\noindent where $U_i$ is the velocity vector, $x_i$, are the coordinate axes such that $i=1,2,3$ are aligned with the streamwise, spanwise, and vertical directions, respectively, $t$ is time, $\langle \cdot \rangle$ represents a spatial plane-average in the streamwise and spanwise directions, $\langle \overline{U}_i\rangle$ is the time- and plane-averaged velocity component, $\widetilde{U}_i$ is the dispersive velocity component, and $u^{\prime}_i$ is the turbulent velocity component. 
In Eq. \ref{eq:velocity_Breakdown}, the repeating index $i$ does not imply tensorial summation.
Expanding the triple decomposition of the velocity vector using the quadratic non-linearity in the incompressible Navier-Stokes momentum equation of the form $U_i U_j$ (dropping the functional notation for simplicity) gives rise to additional stress terms when compared to the conventional Reynolds-decomposition for the time- and plane-averaged quantity $\langle \overline{U_i U_j} \rangle$ given by,

\begin{equation} \label{eq:dispersive_stress}
    \langle \overline{U_i U_j} \rangle = \langle \overline{U}_i \rangle \langle \overline{U}_j \rangle + \langle \overline{\widetilde{U}_i \widetilde{U}_j} \rangle + \langle \overline{u_i'u_j'} \rangle,
\end{equation}

\noindent where the second term on the right-hand side of the above equation is the dispersive stress and the last term is the conventionally defined Reynolds stress. 
A large section of the literature is dedicated to understanding and quantifying the role of dispersive stress in both the rough-wall and canopy flow literature \citep{Wangsawijaya2020,Thakkar2017,Etminan2017,Yuan2014,Nepf2012,Flack2010,Jimenez2004,Perry1969}.
However, understanding the effect of complex multi-scale roughness and its effects has received limited attention, mainly due to the relative complexity required to measure and/or simulate such flows \citep{Anderson2011}. 
As this extra term acts to increase the overall drag experienced by the flow when compared to the canonical turbulent flow in the absence of such a roughness canopy, better understanding and quantifying the effect of this stress term for morphologically complex roughness canopies is important. 
This is motivated by the fact that such turbulent canopies are often observed to take morphologically complex spatial arrangements, thus inducing a considerable heterogeneity in the local hydrodynamic response \citep{Mullarney2017,Blanco2001}

\begin{figure}[hbpt!]
    \centering
    \includegraphics[width=1\linewidth,trim={0cm 2.0cm 0cm 2.5cm},clip]{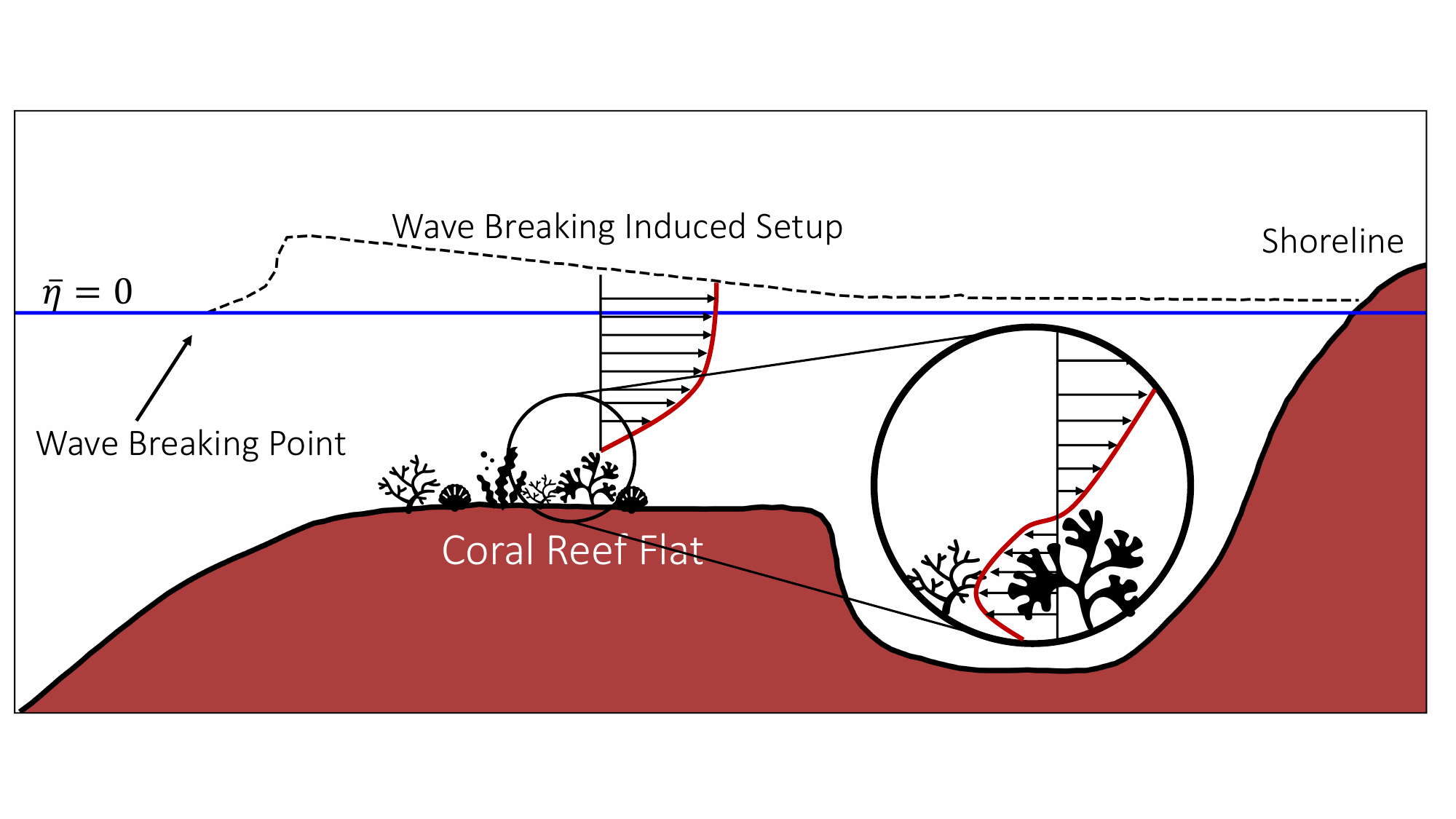}
    \caption{Idealised cross-shore profile depicting the coral reef flat close to the shoreline with the reef profile and wave setup adapted from \cite{Taebi2011}. The red solid line marks the wave setup-induced mean flow on the reef flat over coral roughness.}
    \label{fig:reef_profile}
\end{figure}

Coral reefs in shallow coastal environments are vital ecological organisms that can be characterised as morphologically complex roughness canopies. 
In shallow coastal environments, coral reefs can be observed to inhabit the reef flat regions that can be dominated by turbulent mean flow as a consequence of the wave setup \citep{Taebi2011,Lindhart2021} as sketched in Fig. \ref{fig:reef_profile}. 
As illustrated in the cartoon presented in Fig. \ref{fig:reef_profile}, the wave-induced mean flow over the coral roughness gives rise to non-trivial in-canopy flow response that can have a large impact on the hydrodynamics therein \citep{Gourlay1996,Hench2008,Taebi2011,Hamilton2024}.
In the context of coral reefs, morphological complexity refers to their spatial arrangement and distribution in species and type \citep{Huston1985}. 
More recently, \cite{Hamilton2024} demonstrated the similarities between simple coral roughness types with identical canopy height against a cylindrical and branching roughness canopies. 
However, the effect of morphologically complex coral reef geometries on the flow has not been extensively explored. 
Since the heterogeneity in the coral morphology can induce significant local hydrodynamic effects, in this work, we systematically study the impact of regular simple coral canopies comprising cylinders, branching and massive coral types, and a stochastically generated coral canopy. 
This aims to understand the similarities and differences in the hydrodynamic response at high-flow Reynolds numbers with a special focus on the roughness-induced flow heterogeneity and in-canopy flow characteristics. 
In the following sections, we first detail the governing equations and computational methods used to simulate flow around the complex roughness, which is followed by a detailed analysis of the results and a discussion section. 
Finally, the concluding remarks are presented in the last section.

\section{Computational Methods and Simulation Parameters}

We solve the non-dimensional form of the incompressible Navier-Stokes momentum equations given by

\begin{equation} \label{eq:nondim-Momentum}
    \partial_{t^+} u_i^+ + \partial_{j^+} u_j^+ u_i^+ = \partial_{i^+} p^+ + \frac{1}{Re_{\tau}} \partial_{j^+} \partial_{j^+} u_i^+ + \Pi_c \delta_{i1} + F_{\text{IBM}},
\end{equation}

\noindent and the incompressible continuity equation is given by

\begin{equation}
    \partial_{i^+} u_i^+ = 0,
\end{equation}

\noindent where $t^+$ is time, $x_i^+$ is the coordinate vector, $u_i^+$ is the velocity vector, $p^+$ is the pressure, $Re_{\tau} \equiv u_{\tau} H / \nu$ is the friction velocity based Reynolds number, $u_{\tau}$ is the friction velocity, $H$ is the height of the channel, $\nu$ is the kinematic viscosity of the fluid, $\Pi_c = 1.0$ is the driving pressure gradient, $\delta_{ij}$ is the Kronecker delta, and $F_{\text{IBM}}$ is the immersed boundary force, respectively. The coordinate vector is oriented such that $x_1^+$, $x_2^+$, and $x_3^+$ correspond to the streamwise, spanwise, and vertical directions, respectively.
In Eq. \ref{eq:nondim-Momentum}, the non-dimensionalisation (terms with a $+$ super-script) is carried out using a reference velocity $u_{\tau}$, reference length scale $H$, and pressure scaling $p^+ \equiv p/\rho u_{\tau}^2$, where $\rho$ is the density of the fluid. 
The differential form of the governing equations is spatially discretised using the second-order accurate finite-difference method combined with the Runge-Kutta three-step time integration scheme based on the fractional step algorithm \citep{KimMoin1985,FerzigerPericStreet2019}. 
All the terms are discretized explicitly, and the time-step is adjusted such that the Courant–Friedrichs–Lewy (CFL) number of 0.9 is not exceeded. 
To achieve scalability, the computational domain is decomposed using a spanwise slab method as detailed in \cite{LiLaizet2010} and parallelised using the message-passing interface (MPI) library \citep{Gabrieletal2004}.
The complex roughness is introduced using the \textit{GenSDF} software that accurately masks the roughness using an efficient geometry-local signed-distance-field (SDF) algorithm \citep{Patil2025}, through which the fluid and solid regions of the flow are identified, and a volume penalising immersed-boundary method is employed \citep{Scotti2006}. 
The solid region can be identified as the grid points where the SDF has a negative value, i.e., the grid points lie within the watertight manifold geometry. 

\begin{figure}[hbpt!]
    \centering
    \includegraphics[width=1\linewidth]{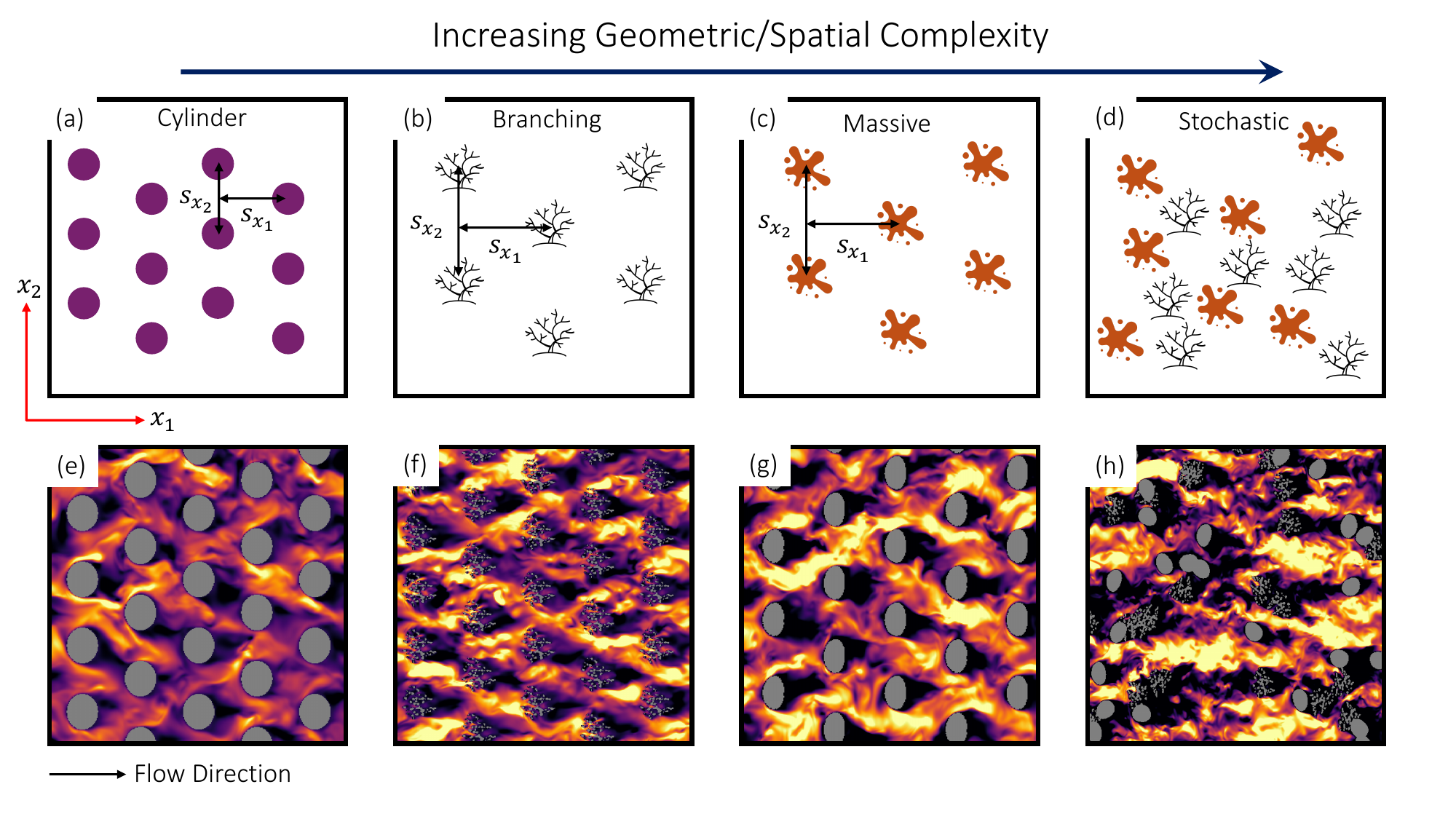}
    \caption{Panels (a-d) detail sketches for the various flow cases considered in this study, with the horizontal and vertical arrows marking the streamwise and spanwise spacing between the individual geometries. Panels (e-h) depict the streamwise velocity fluctuation ($u_1^{\prime}$), where the grey regions denote the solid region. The colours represent the streamwise velocity; darker (black) colours represent slow velocity, and brighter (yellow) colours represent fast velocity. The red arrows mark the directions for the coordinate axes corresponding to the streamwise and spanwise directions, respectively.}
    \label{fig:figure1}
\end{figure}

In this work, all the simulations are run at $Re_{\tau} = 1000$ and a fixed height of the individual roughness element ($k_c = 0.1~$m) in a channel with dimensions $L_{x_1} \times L_{x_2} \times L_{x_3} = 60 k_c \times 30 k_c \times 10 k_c$. 
The channel geometry is discretized using $N_{x_1} \times N_{x_2} \times N_{x_3} = 2000 \times 1500 \times 400$ grid cells, respectively, where $x_3 < 0.15$ is discretized using $175$ grid points resulting in a constant grid $\Delta x_3^+ = 0.86$, above which  hyperbolic tangent grid stretching is used with $\Delta x_{3,\text{max}}^+ = 6.8$ at the top of the channel. 
The channel is periodic in the streamwise and spanwise directions, while a no-slip boundary condition is applied at the bottom wall, and a free-slip boundary condition is applied at the top wall.
This geometry is designed to emulate the key features of the coral flat detailed in Fig. \ref{fig:reef_profile}, which comprises a relatively flat topography and presence of a coral canopy.
We consider four cases in total as detailed in Fig.~\ref{fig:figure1}, with the streamwise and spanwise spacing for the staggered cases $s_{x_1} = s_{x_2} = 2k_c$ and for the stochastic case, the total number of corals in the simulation domain are kept identical to the ones simulated in the staggered arrangement. 
Here, the stochastic spatial configuration is achieved through a two-stage process: (a) first, a list of $x_1$ and $x_2$ coordinates are sampled from a uniform distribution and used as the centroid of the translation vector; (b) next, the triangulated coral geometry is translated about its centroid from the origin at $x_o \equiv (0,0,0)$ to the translation point $x_t \equiv (x_1,x_2,0)$ along with a rotation about the vertical axis where the rotation angle is sampled from a uniform distribution with a range [$0^{\circ}$,$360^{\circ}$]. 
The \textit{Acropora Formosa} \citep{Smithsonian_Acropora} and \textit{Pseudodiploria Strigosa} \citep{Smithsonian_Pseudodiploria} species of branching and massive coral types, respectively, are used to generate the stochastic coral reef as shown in Fig.~\ref{fig:figure1}h.
This method to generate a stochastic coral reef bed is relatively simple and yields a sufficiently complex morphology of the coral reef at a minimal computational cost. 
Advanced techniques through the use of data-driven methods \citep{Crocker2024} do exist; however, the lack of species-classified coral reef data and inherent limitations associated with the 3D modelling \citep{brouwer2024} limit their application at scale; thus, a relatively simple stochastic coral reef generator was chosen in this work.

All simulations are run using the SURF-Snellius high-performance computing cluster on the \textit{Genoa} partition hosting 192 CPUs with 336 GiB of memory per node. 
Each simulation used 500 CPUs over three \textit{Genoa} nodes and required a total of $160 \times 10^3$ CPU-hours to simulate $30$ eddy-turn-overs ($T_{\epsilon} \equiv H/u_{\tau}$). 
To reduce the initial spin-up time associated with the transient from flow transition to a statistically stationary flow condition, a synthetically generated three-dimensional flow field was generated using \textit{GenIC} \citep{PatilGarciaSanchez2025}. 
To generate this three-dimensional flow field, \textit{GenIC} requires the vertical profiles for the time- and plane-averaged velocity and the Reynolds stress tensor. 
These input profiles are obtained from periodic simulations carried out by \cite{Castro2006} and are rescaled to the target value of $Re_{\tau} = 1000$ using inner scaling. 
This effectively generates a suitable initial condition that transitions and achieves a statistically stationary flow state within $5T_{\epsilon}$. 
In the discussion that follows, all the time- and plane-averages are carried out over the last $25T_{\epsilon}$ after an initial transient of $5T_{\epsilon}$.

\section{Double-averaged mean flow statistics}

\begin{figure}[hbpt!]
    \centering
    \includegraphics[width=1\linewidth,trim={0cm 1.75cm 0cm 2cm},clip]{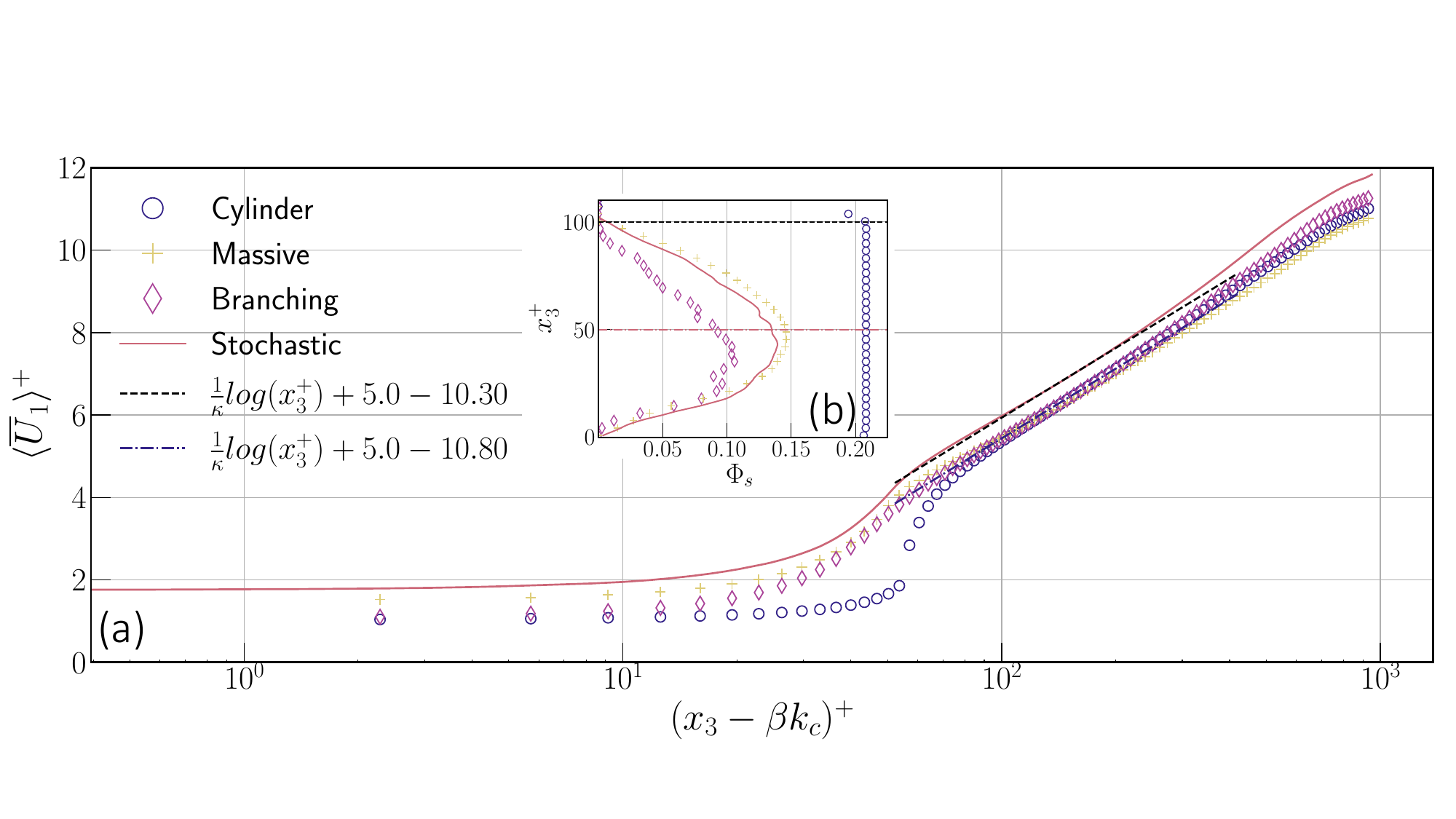}
    \caption{Double-averaged velocity profile for the various cases considered in this work. For the velocity profile, every fourth data marker is shown for all the cases. The inset in the middle shows the corresponding solid fraction ($\Phi_s$) as a function of the vertical distance from the wall ($x_3$) for the various cases.}
    \label{fig:figure2}
\end{figure}

The solidity fraction ($\Phi_s$) is given by

\begin{equation}
    \Phi_s \left( x_3 \right) = \frac{1}{L_{x_1} L_{x_2}} \sum_{i=1}^{i=N_1} \sum_{j=1}^{j=N_2} \phi_{i,j} \Delta x_{1}^{i} \Delta x_2^{j}, 
\end{equation}

\noindent where $L_{x_1}$ and $L_{x_2}$ is the length of the domain in the streamwise and spanwise directions, respectively, $\phi_{i,j}$ is the binary solid indicator field (1 for solid and 0 for fluid phase), and $\Delta x_1^{i}$ and $\Delta x_2^{j}$ are the grid spacing in the streamwise and spanwise directions, respectively.  
The primary role of the dispersive stresses is to introduce additional drag, which can be observed in the plane- and time-averaged (hereafter double-averaged) velocity profile as presented in Fig.~\ref{fig:figure2}. 
The displacement height for all the cases discussed in this work is obtained by setting $\beta = 0.5$, corresponding with the peak of $\Phi_s$ in the vertical direction for the non-cylindrical roughness cases. 
The von K\'{a}rm\'{a}n constant is taken to be $\kappa = 0.41$. 
Comparing the cylinder, branching, and massive cases, significant differences in the double-averaged profiles can be observed throughout the velocity profile. 

Close to the wall, the velocity profile for the cylinder case is observed to be affected the most as the velocity recovery is relatively slow when compared to the other two cases (i.e., branching and massive) arranged in a staggered fashion. 
This is not surprising since, for the cylinder case, there are no vertical variations in $\Phi_s$; thus, the relative frontal area is larger compared to the other cases. 
Consequently, the flow experiences a larger form drag penalty when compared with the branching and the massive cases that are arranged similarly. 
This effect can be readily seen in Fig.~\ref{fig:figure1} panels e, f, and g, respectively, where the case with massive corals is observed to exhibit relatively larger velocity followed by the branching corals, and cylindrical canopy exhibits the lowest velocity amongst the three cases albeit the instantaneous nature of the snapshot presented therein. 
The double-averaged velocity profiles do not seem to have a unique trend as a function of $\Phi_s$ when the staggered cases are considered. 
Specifically, the peak value of $\Phi_s$ for the branching coral case is relatively smaller than that of the massive coral case; however, the impact of the corals on the double-averaged velocity is observed to be relatively larger for the branching coral case as it exhibits a lower velocity when compared to the massive coral case. 
The stochastic coral case exhibits a relatively larger velocity magnitude in comparison to all the cases discussed in this paper, despite having a solid fraction similar to the massive coral case (See Fig.~\ref{fig:figure1}e-h and Fig.~\ref{fig:figure2}b).
For all the cases discussed in this work, a strong outer layer imprint is observed above the inertial range indicated by the velocity profile above $(x_3 - \beta k_c)^+ \gtrsim 400.0$.
This outer layer is relatively consistent for the Massive, Cylinder, and Branching cases, respectively, in decreasing order of influence, while the inertial range is observed to be relatively similar for these three cases.
Apart from these observed differences, the staggered coral cases exhibit a relatively consistent trend, as seen by the clustering of the velocity profile in the inertial range, barring a weak dependence on the underlying coral type.

\section{Double-averaged Root Mean Square Velocity and Reynolds stress profiles}

\begin{figure}[hbtp!]
    \centering   
    \includegraphics[width=\linewidth]{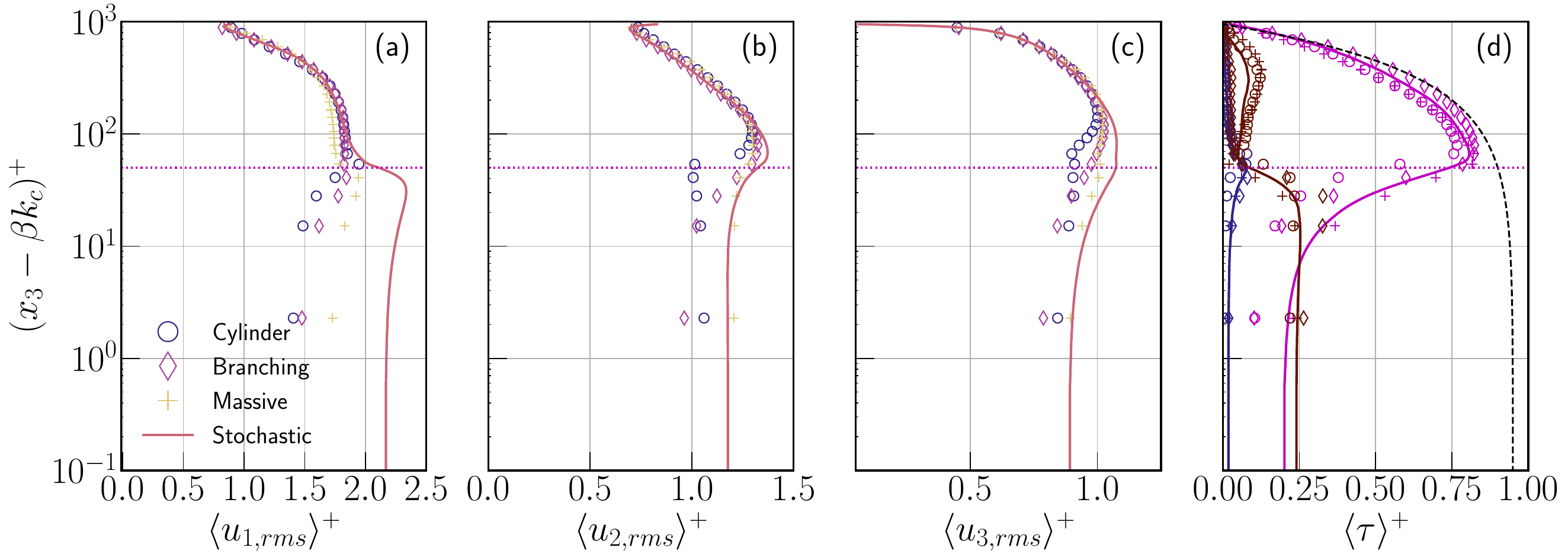}
    \caption{(a-c) Double-averaged rms velocity profiles for the streamwise, spanwise, and vertical velocity components as a function of the vertical distance from the wall. (d) Double-averaged viscous stress (blue), Reynolds stress (magenta), and dispersive stress (brown). The black-dashed line marks the linear stress profile. The horizontal dotted magenta line in all the panels marks the top of the coral geometry (i.e., $k_c^+$). In all the panels, one in every fifteenth data marker is shown.}
    \label{fig:figure3}
\end{figure}

\begin{figure}[hbtp!]
    \centering   
    \includegraphics[width=\linewidth]{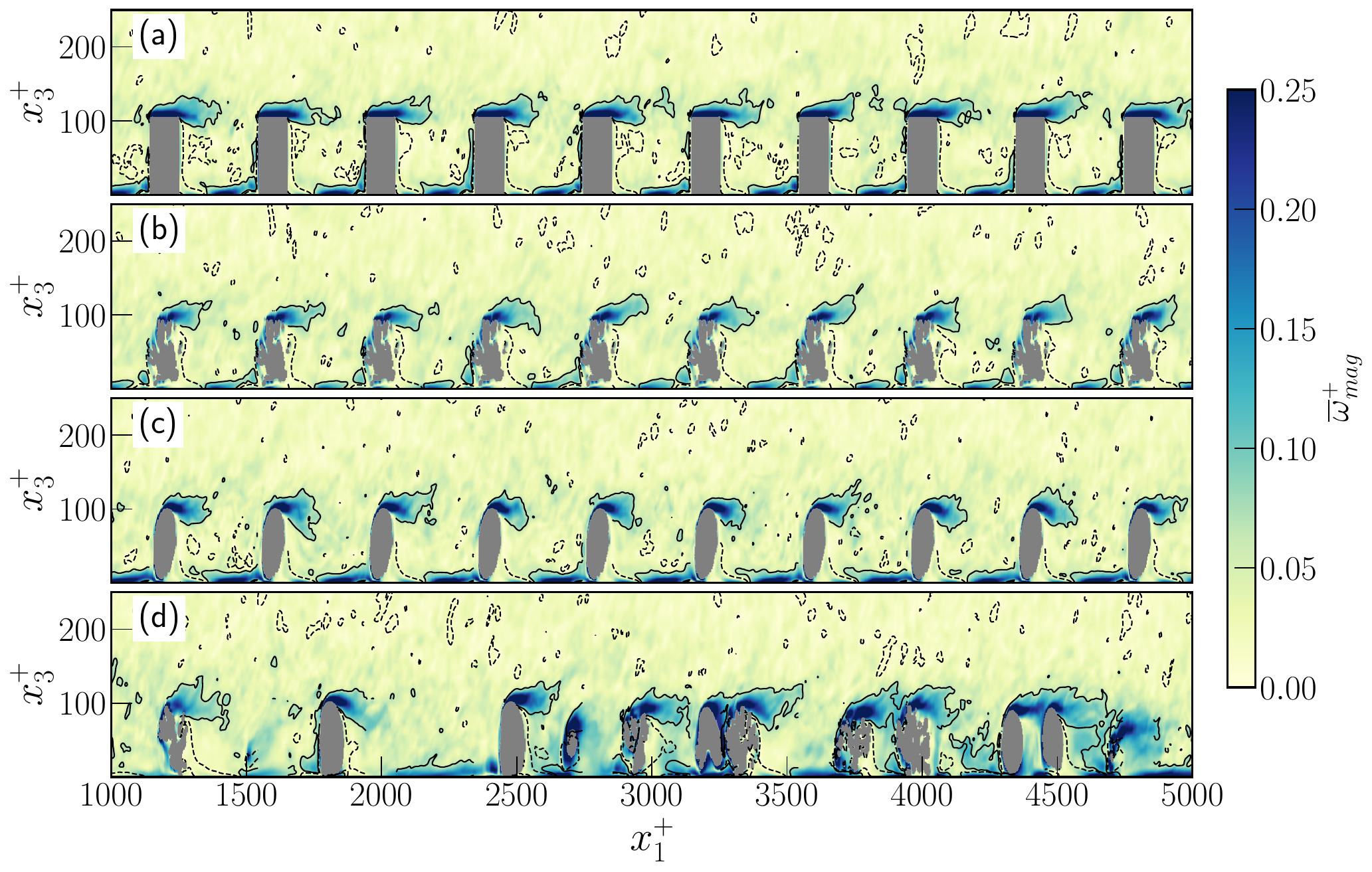}
    \caption{Comparison of the time-averaged vorticity magnitude ($\overline{\omega}_{mag}^+$) at a slice located at $x_2^+ = 2050$. The grey area marks the coral roughness. The contours overlaid on the pseudo-colour map denote the time-averaged vorticity about the spanwise axis (i.e., $\overline{\omega}_{2}^+$), where the dashed black line corresponds to a value of $\overline{\omega}_{2}^+ = 0.0$ and the solid black line corresponds to a value of $\overline{\omega}_{2}^+ = 0.0625$. Vorticity magnitude on all the panels is non-dimensionalised using $\nu/u_{\tau}^2$.} 
    \label{fig:vorticity}
\end{figure}

To further understand the differences observed in the various coral cases, we compare the double-averaged root-mean-squared (rms) velocity profiles along with the double-averaged viscous stress, Reynolds stress, and dispersive stress in Fig.~\ref{fig:figure3}. 
As shown in Fig.~\ref{fig:figure3}a, a systematic difference can be observed when comparing the four cases. 
Below the top of the coral canopy (i.e., $k_c$), the cylinder case is observed to exhibit the lowest magnitude of $\langle u_{1,\text{rms}} \rangle^+$, followed by the branching, massive, and stochastic cases, in ascending order, respectively.
Above the coral canopy, the $\langle u_{1,\text{rms}} \rangle^+$ profiles for the stochastic, branching, and the cylinder cases are found to agree well, while the massive case is observed to have a relatively lower magnitude until $(x_3 - \beta k_c)^+ \sim 300$, beyond which all the profiles are identical using inner scaling parameters.
The stochastic case has a peculiar peak in the $\langle u_{1,\text{rms}} \rangle^+$ profile close to the top of the coral canopy, primarily due to the heterogeneous nature of the coral morphology.
While it is well understood that the location of the peak $\langle u_{1,\text{rms}} \rangle^+$ is sensitive to the Reynolds number \citep{Ma2021}, the differences observed in this case are not a consequence of changing Reynolds numbers but due to the differences in the coral morphology.
This is clearly illustrated in Fig. \ref{fig:vorticity}, which compares the time-averaged vorticity magnitude for the various cases considered in this study.
For the cylinder, branching, and the massive cases (Fig. \ref{fig:vorticity}a-c), positive vorticity is distributed primarily very close to the wall in the upstream region of the roughness and at the crest of the roughness elements, indicative of the strong shear layer.
The $\overline{\omega}_{mag}^+$ peak upstream of the roughness elements can be attributed to the shear at the wall as well as the horseshoe or saddle type vortex typically observed upstream of bluff bodies \citep{Sumer1997,Escauriaza2011}.
There also exists a systematic difference in the overall structural orientation of the peak vorticity at the coral roughness crests between the four cases as illustrated in Fig. \ref{fig:vorticity}.
For the cylinder, branching, and massive cases, the peak vorticity magnitude is located at the crest of the coral roughness, indicative of the shear layer that develops in this region. 
While for the stochastic case, the peak vorticity is distributed over the height of the canopy, unlike the other cases.
This explains the relatively large streamwise rms velocity peak observed for the stochastic case when compared to the rest of the cases. 

For the $\langle u_{2,\text{rms}} \rangle^+$ and $\langle u_{3,\text{rms}} \rangle^+$, a similarly consistent observation can be made for the stochastic case, which suffers from high levels of turbulence throughout the canopy.
For all the rms velocity components, there is good agreement between the various cases away from the wall, suggesting that most of the significant changes are contained within the roughness canopy \citep{Ghisalberti2002,Finnigan2009}. 
Comparison of the $\langle u_{3,\text{rms}} \rangle^+$ profiles shows that, in all cases, the results converge consistently beyond $(x_3 - \beta k_c)^+ \approx 200$. This indicates that, from a phenomenological perspective, there are no discernible differences between the cases sufficiently far above the canopy when the rms velocity components are considered. 
While this is true for the rms velocity components, the double-averaged stress profiles do not reflect this unified behaviour.
Starting with the smallest magnitude term, the viscous stress term is consistently small in magnitude across all the cases, which is expected, given the bluff nature of the coral roughness, in agreement with the experimental findings of \cite{Yang2015}.
The dispersive stress profiles are observed to have the smallest magnitude for the branching case throughout the vertical profiles, while the massive and the cylinder cases exhibit relatively higher magnitudes.
Within the coral canopy, the dispersive stress has a peak value of $\sim 0.25$ when non-dimensionalised using the friction velocity (obtained from $\Pi_c$ in Eq. \ref{eq:nondim-Momentum}), and there after stays below this threshold for all the cases in agreement with previous numerical observations \cite{Yuan2014,Yang2015,Brereton2021}.
The Reynolds stress profiles are observed to have an overall similar trend, but are seen to be sensitive to the underlying coral roughness.
Specifically, the branching case shows relatively large magnitudes of Reynolds stress, followed by the massive, stochastic, and cylinder cases, respectively.
At the centre of the channel (inertial range), the Reynolds stress profiles are observed to deviate from the conventional flat wall counters with a simultaneous increased contribution by the dispersive stress terms as observed in \cite{Yuan2014,Yang2015,Brereton2021}.
These observations correlated with the distribution of the vorticity as shown in Fig. \ref{fig:vorticity} seem to suggest that while the rms velocity components are observed to collapse well far from the wall, both the mean velocity and the Reynolds stress profiles are observed to be sensitive to the underlying coral geometry.

\section{Flow within the coral canopy}

\begin{sidewaysfigure}[hbpt!]
    \includegraphics[width=0.95\linewidth,angle=0,trim={0.2cm 3cm 0.5cm 3cm},clip]{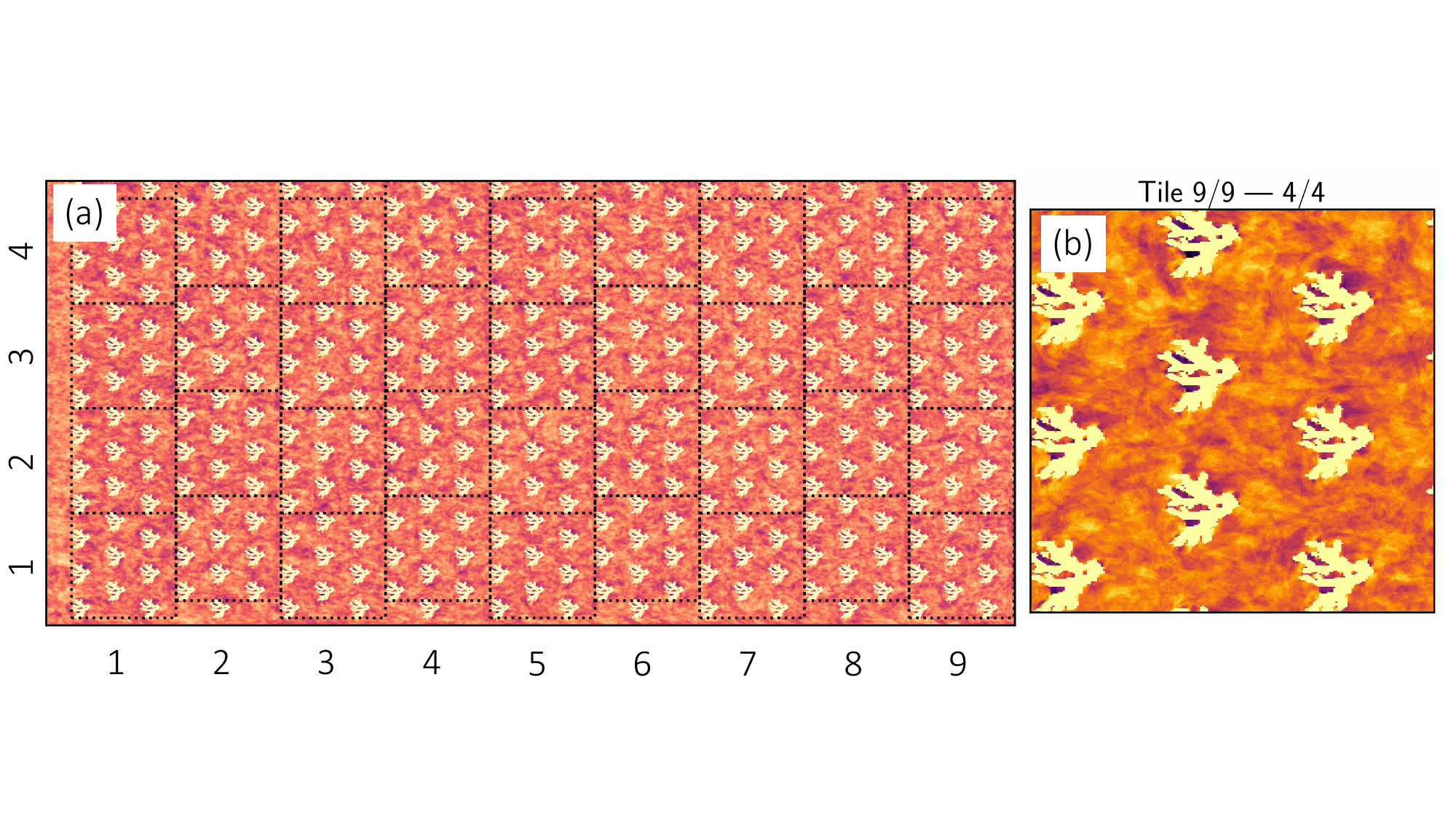}
    \caption{(a) Top view of the symmetric tiles marked with black dashed lines (tiles of relatively larger size, for illustration) for the branching case with the colourmap representing the contribution of Reynolds stress through the 1$^{st}$ quadrant interactions. The numbers along the x-axis and the y-axis correspond to the columns and rows for each tile, respectively. (b) Colourmap corresponding to the 1$^{st}$ quadrant contribution to the Reynolds stress for the last tile. The mean flow in both the panels is going from left to right with the x and y axes of the panels aligned along the streamwise and the spanwise directions, respectively, while the slice is taken at $x_3^+ = 89$.}
    \label{fig:sampling}
\end{sidewaysfigure}

As detailed in the double-averaged statistics and the time-averaged contours of vorticity, the flow within the coral canopy is observed to be sensitive to the type of coral roughness.
While Fig. \ref{fig:figure3}a-c illustrate some agreement outside the coral canopy, the differences observed in the shear stress profiles and the vorticity magnitude in Figs. \ref{fig:figure3}d and \ref{fig:vorticity}, respectively, demand a closer look at the time-averaged in-canopy flow.
For the staggered cases, the computational domain can be further reduced by leveraging the spatial symmetry that exists thanks to the staggered arrangement of the coral roughness, however, such a sub-domain reduction is not possible for the stochastic case due to the inherent heterogeneity in the coral arrangements.
The symmetric tiles are illustrated in Fig. \ref{fig:sampling} and are used to average the mean flow and other flow statistics presented for the staggered cases.

\begin{sidewaysfigure}[hbpt!]
    \includegraphics[width=0.55\linewidth,angle=270]{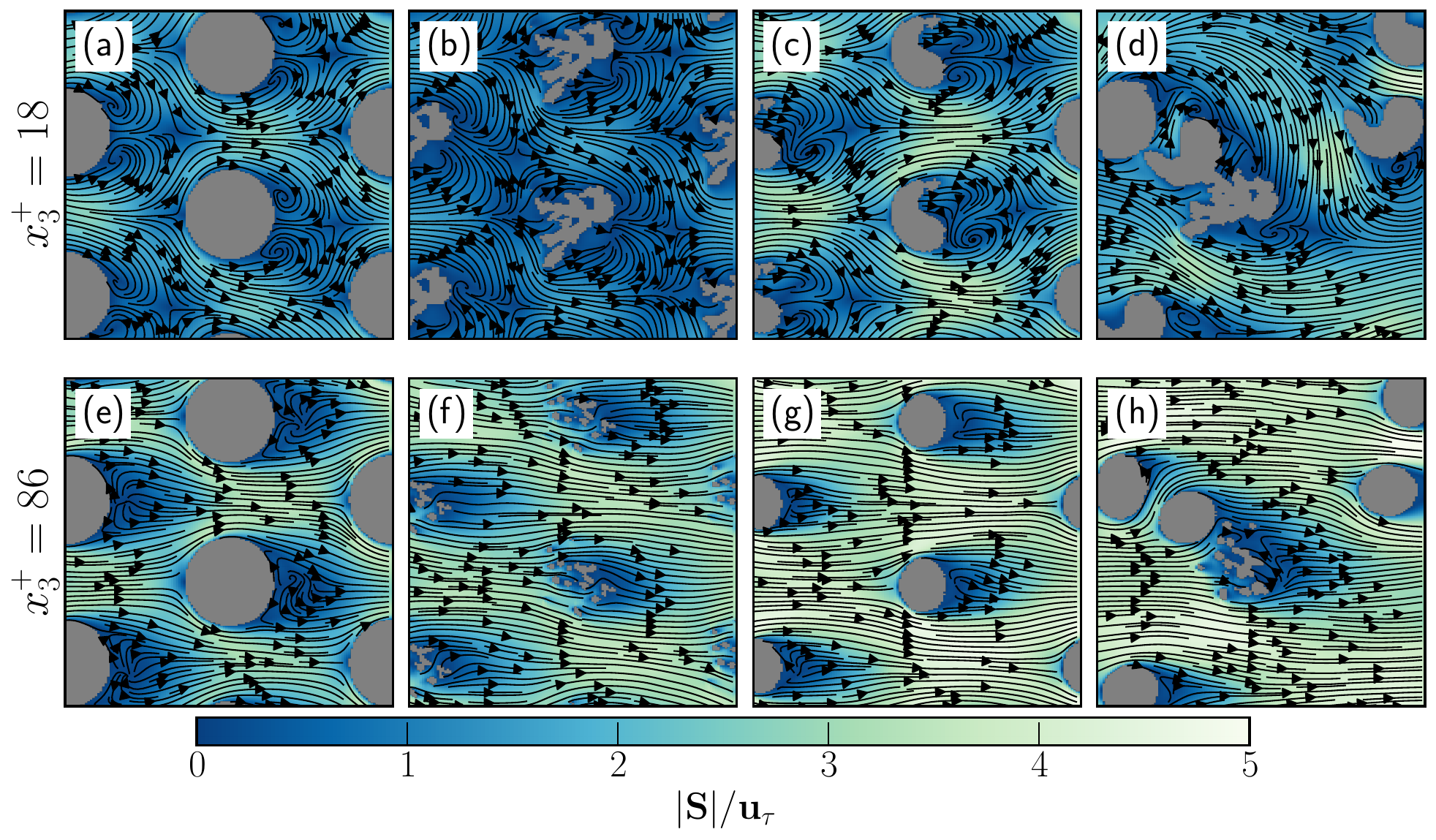}
    \caption{Comparison of the in-canopy flow for the various cases considered in this paper. Panels (a)-(d) show the horizontal slice at $x_3^+ = 18$, and panels (e)-(h) show identical horizontal spatial location as panels (a)-(d) at $x_3^+ = 86$. The pseudo colour marks the velocity magnitude ($\mathbf{S}$), grey colour on all the panels denotes the coral roughness, and solid black lines with arrows mark the time-averaged streamlines. }
    \label{fig:symmetry_streamlines}
\end{sidewaysfigure}

Leveraging this spatial symmetry for the staggered arrangements, Fig. \ref{fig:symmetry_streamlines} shows the time-averaged streamlines overlaid on the pseudo-colour map denoting the velocity magnitude within the coral canopy.
For the cylinder case, since the planform-area density does not change as a function of height, the time-averaged streamlines appear to be similar across the height of the coral canopy with some minor differences.
Closer to the bottom wall, the typically observed velocity magnitude is lower compared to its higher counterpart, as illustrated in Figs. \ref{fig:symmetry_streamlines}a and e, where the wakes behind the cylindrical roughness are relatively symmetric as a consequence of the lower velocity magnitude.
For the branching case, close to the wall (Fig. \ref{fig:symmetry_streamlines}b), the planform area is relatively larger and thus experiences a relatively stronger wake when compared to the velocity and streamlines higher within the canopy (Fig. \ref{fig:figure3}f).
The branching coral geometry also exhibits an asymmetry about the x-axis, which gives rise to a relatively complex wake as a function of distance away from the bottom wall (Figs. \ref{fig:symmetry_streamlines}b and f).
The massive case also exhibits a similar difference as seen for the branching case, where the canopy flow close to the wall is relatively different from the flow further away.
This is primarily an imprint of the changing planform area of the coral roughness as seen in Figs. \ref{fig:symmetry_streamlines}c and g.
For the stochastic case shown in Figs. \ref{fig:symmetry_streamlines}d and h, the relatively complex spatial arrangement introduces a stronger heterogeneity in the in-canopy flow throughout the canopy when compared to the other three cases.
This is evident from the differences observed between the two in-canopy heights depicted in Fig. \ref{fig:symmetry_streamlines}, where close to the wall (Fig. \ref{fig:symmetry_streamlines}d) the time-averaged flow is observed to have a larger spanwise component in the top centre while the bottom left part of the depicted region shows a stronger streamwise flow direction. 
Further away from the wall, the stochastic case shows a relatively stronger streamwise flow (Fig. \ref{fig:symmetry_streamlines}h) as the planform area changes significantly with increasing distance from the wall.
Overall, these observations suggest that the in-canopy flow features are relatively sensitive to the coral geometry type and significantly affect the mean flow as depicted in Figs. \ref{fig:vorticity} and \ref{fig:symmetry_streamlines}.

\section{Stress within the coral canopy}
\label{sec:stres_within_canopy}

\subsection{Staggered Cases}
\label{sec:staggered_incanopy_stress}

\begin{figure}[hbpt!]
    \includegraphics[width=0.8\linewidth,trim={0.5cm 0cm 0.3cm 0cm},clip]{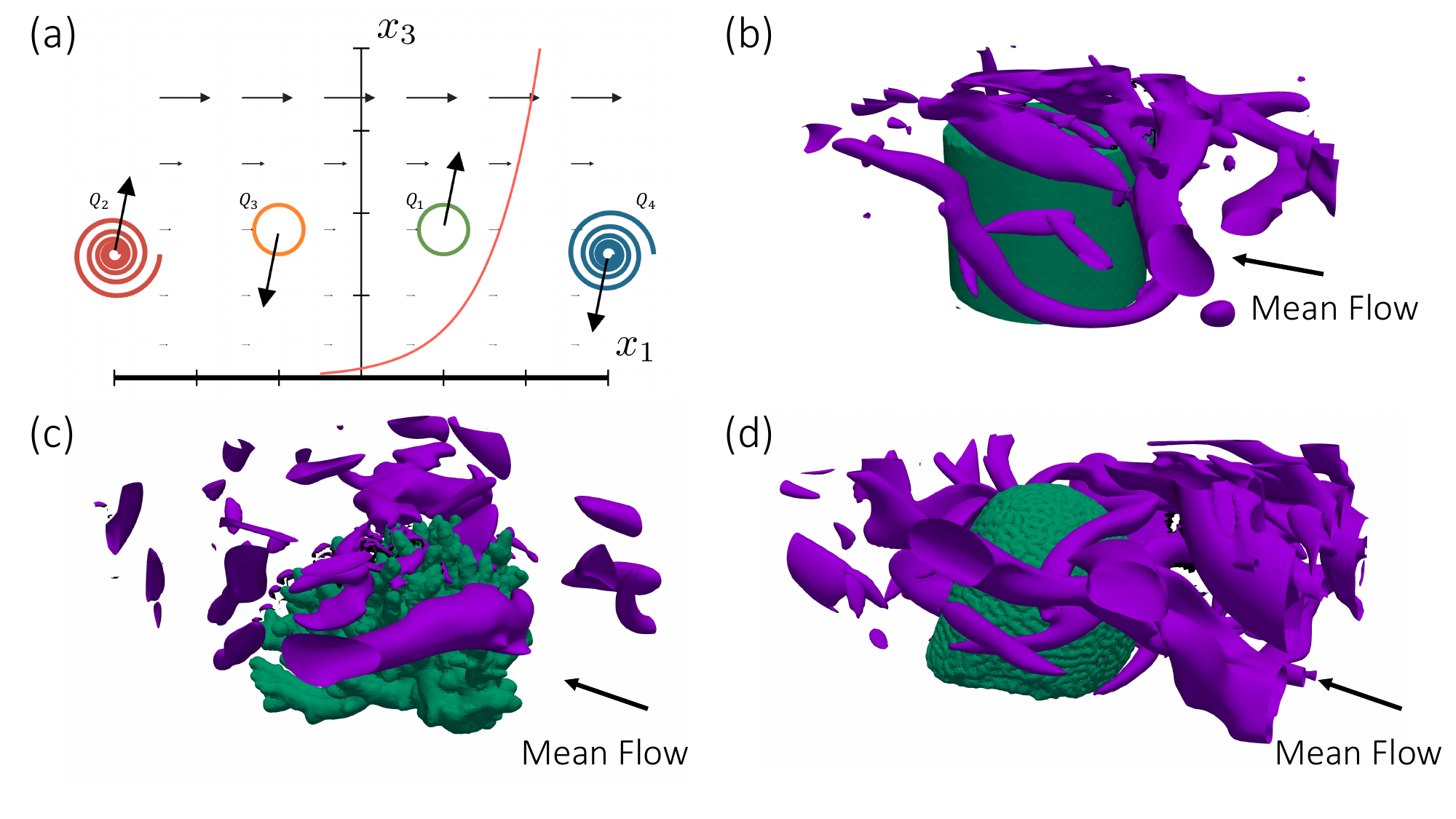}
    \caption{(a) Schematic representing the four primary interactions contributing to the Reynolds stress (i.e., the time-averaged value $\overline{u_1^{\prime} u_3^{\prime}}$). Circles with arrows represent wall-ward ($Q_3$) and outward ($Q_1$) interactions, while the spirals with arrows represent ejections ($Q_2$) and sweeps ($Q_4$), respectively. The red solid line marks the logarithmic velocity profile while the black arrows indicate the background flow. The sketch assumes a homogenous spanwise direction ($x_2$) going into the page. (b-d) Q-criterion visualised for the staggered cases at an arbitrary instance illustrating the \textit{necklace}-like vortex upstream of the coral roughness coloured in purple, while the teal colour represents the coral roughness. Panels b, c, and d correspond to the cylinder, branching, and the massive cases, respectively.}
    \label{fig:quadrant_interactions}
\end{figure}

\begin{figure}[hbpt!]
    \includegraphics[width=1\linewidth]{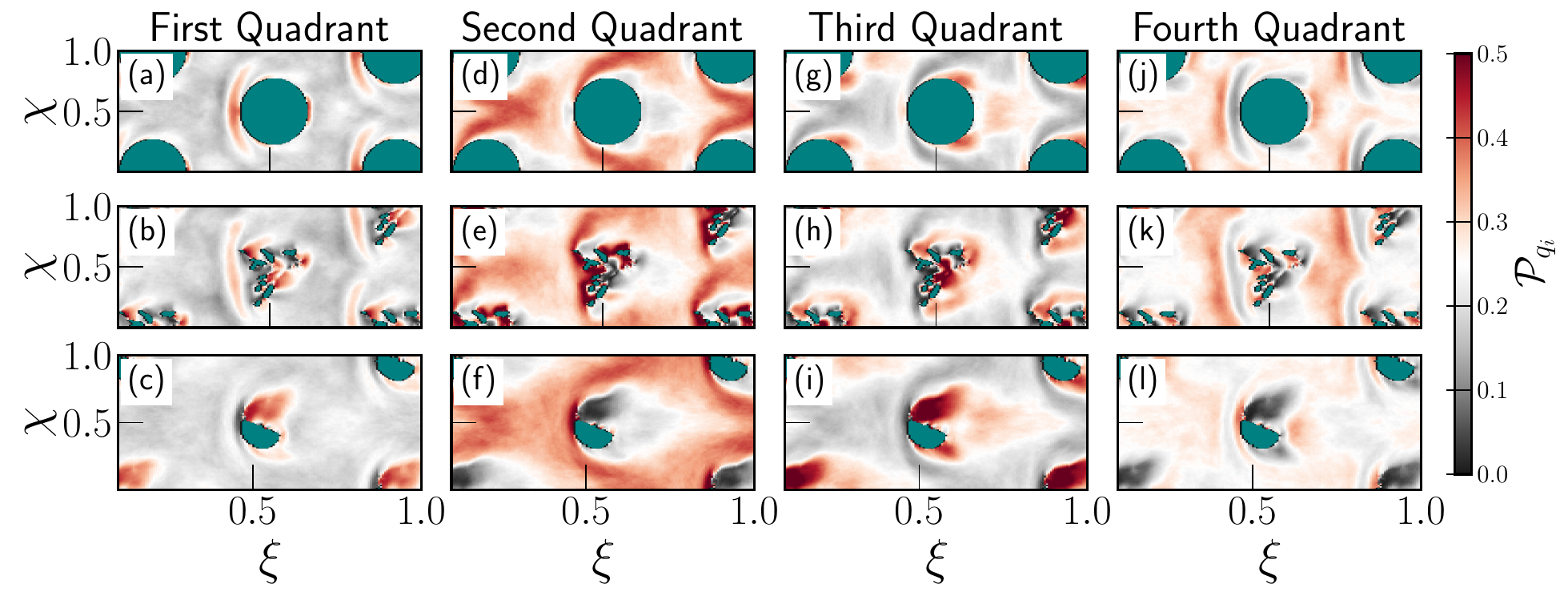}
    \caption{Comparison of the probability for quadrant contribution to the time-averaged Reynolds stress for the staggered cases within the canopy, close to the wall. Each column marks the contribution of the quadrant, while each row marks the cylinder, branching, and massive case, respectively. The pseudo-colour marks the probability for the respective quadrant as a function of space, with red regions denoting a larger probability and black regions denoting a lower probability of occurrence.}
    \label{fig:quadrant_slices}
\end{figure}

In the previous section, the in-canopy mean flow was presented in extensive detail, where the stochastic case was observed to exhibit a distinct in-canopy and above-canopy response when compared to the staggered cases. 
To further elucidate this sensitivity to the underlying coral geometry, the changes to the contribution towards the Reynolds stress are discussed in this section. 
Using the spatial symmetry for each repeating tile for the staggered cases (see Fig. \ref{fig:sampling}), the streamwise and spanwise coordinates can be transformed to a tile local coordinate system. 
For each individual tile, the origin is set at the bottom left corner of the tile, and the streamwise coordinate axis is transformed as $\xi \equiv [0,x_1/(4k_c)]$, while the spanwise coordinate axis is transformed as $\chi \equiv [0,x_2/(3k_c)]$.
As sketched in Fig. \ref{fig:quadrant_interactions}, there are four primary interactions, which can be binned based on the quadrants corresponding to the sign of the fluctuating velocity components contributing to the Reynolds stress \citep{Wallace2016}.
To statistically quantify the occurrence of the various types of motions that contribute to the Reynolds stress, as detailed in Fig. \ref{fig:quadrant_interactions}a, the probability of occurrence ($P_{q_i} \left( \xi, \chi\right)$) at each grid point over the symmetric tile (see Fig. \ref{fig:sampling}) is calculated by binning the Reynolds stress term into one of the four bins per sample.
The probability of occurrence ($P_{q_i} \left( \xi, \chi\right)$) as defined in this context aims to characterise the frequency with which the quadrant events occur over the symmetric tiles for the various coral roughness considered in this work.
It is pertinent to note that a relatively larger value of $P_{q_i} \left( \xi, \chi\right)$ only marks the occurrence frequency and not the overall importance of the quadrant contribution towards the average value of the Reynolds stress ($u_1^{\prime} u_3^{\prime}$), as that is determined by the magnitude of the product of the fluctuating quantities.
Consequently, a relatively large occurrence probability does not entail a larger contribution towards the Reynolds stress as the product of the fluctuating quantities may be relatively small, thus affecting the time-averaged value minimally.
While this is true, it is important to understand the spatial extent of these events as the quadrant events are correlated with physical mechanisms that transport momentum wallwards (or away from it), and thus potentially have deeper implications for the in-canopy mixing potential in the presence of coral roughness.
Fig. \ref{fig:quadrant_slices} compares the three staggered cases over these symmetry tiles at $x_3^+ = 10.0$ with the pseudo-colour denoting $P_{q_i} \left( \xi, \chi\right)$ and the teal region marking the coral roughness.
The contribution of the Reynolds stress as a function of the distance from the wall is included in Appendix \ref{sec:appendixA} and will not be discussed here for brevity.

Comparing the first and the fourth quadrant events across the three cases (panels a-c and j-l in Fig. \ref{fig:quadrant_slices}), there is a clearly visible upstream recirculation zone at the base of the coral roughness marked by the relatively stronger $Q_1$ and $Q_4$ events attached to the coral roughness.
This recirculation zone is indicative of the vortex tube that is formed as a consequence of the stagnation point at the base of the coral roughness, thus giving rise to the horseshoe or necklace vortex \citep{Escauriaza2011,Kirkil2015}.
The base vortex was also identified by visualising the Q-criterion \citep{Hunt1988} for a series of time-snapshots that revealed this attached vortex at the upstream base of the coral roughness, with one such snapshot shown in Fig. \ref{fig:quadrant_interactions}b.
Making a similar comparison across the various quadrants (e.g., panels a, d, g, and j) for a given coral roughness type, it is clear to see that $P_{q_2} \left( \xi, \chi\right)$ dominates the spatial extend closely followed by $P_{q_4} \left( \xi, \chi\right)$ and subsequently $P_{q_3} \left( \xi, \chi\right)$ and $P_{q_1} \left( \xi, \chi\right)$, for all the cases presented in Fig. \ref{fig:quadrant_slices}.
These observations suggest that for the staggered cases, there is a prevalence of $Q_1$ and $Q_4$ events localised at the upstream of the coral roughness.
The presence of the upstream \textit{necklace}-type vortex \citep{Kirkil2015} might help explain this trend observed for the staggered cases.

\begin{figure}[hbpt!]
    \includegraphics[width=1\linewidth]{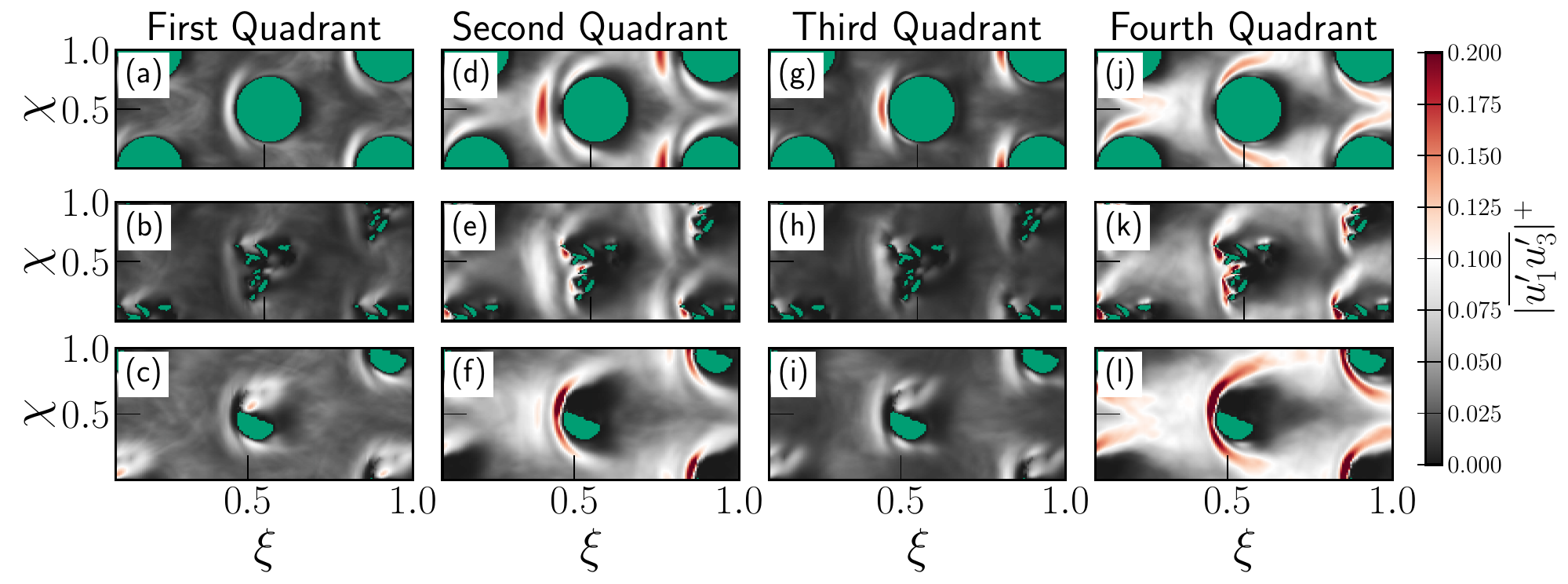}
    \caption{Comparison of the tile- and time-averaged quadrant magnitude for the staggered cases within the canopy, close to the wall. Each column marks the mean value of the quadrant, while each row marks the cylinder, branching, and massive case, respectively.}
    \label{fig:quadrant_mean_value}
\end{figure}

\begin{figure}[hbpt!]
    \includegraphics[width=1\linewidth]{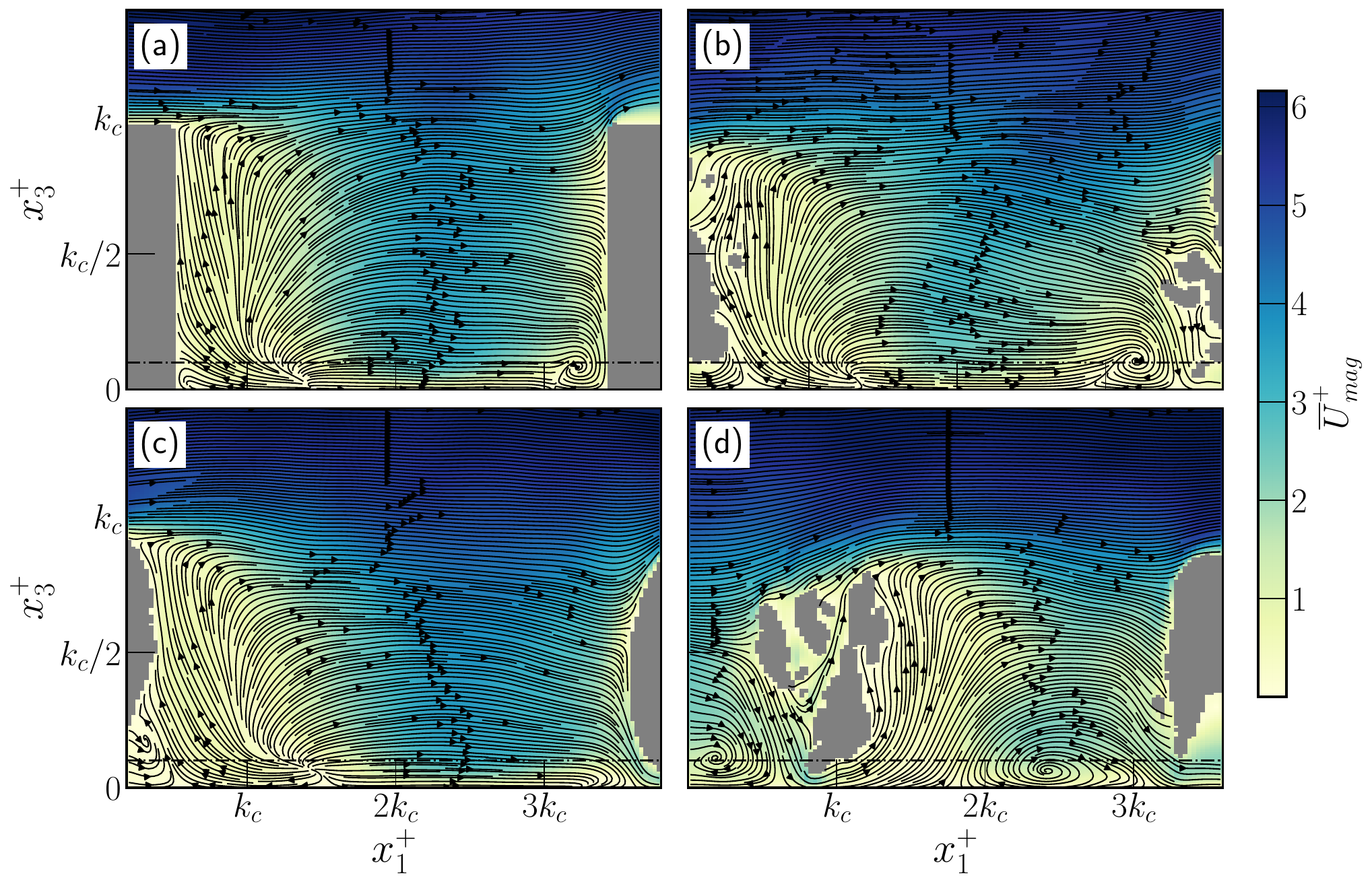}
    \caption{Time-averaged streamlines plotted using the streamwise and the vertical velocity components for the four cases discussed in this work. Panels a-d are marked with the background pseudo-colour illustrating the velocity magnitude. For the stochastic case shown in panel d, a representative portion of the streamwise location is chosen to illustrate the heterogeneity when compared to the staggered cases shown in panels a-c. The horizontal line on each panel corresponds to $x_3^+ = 10$.}
    \label{fig:vertical_streamlines}
\end{figure}

In the preceding sections, the spatial extent of the various quadrant events was discussed, which details the occurrence frequency of these events that can be tied to specific mechanisms of momentum transport \citep{Finnigan2009}.
The spatial distribution of the individual quadrant contributions can be seen in Fig. \ref{fig:quadrant_mean_value} at $x_3^+ = 10$ for the staggered cases.
Contrasting the $Q_1$ events and their peak magnitudes for the three cases (panels a-c in Fig. \ref{fig:quadrant_mean_value}) against the peak value of $P_{q_1} \left( \xi, \chi\right)$ shown in Fig. \ref{fig:quadrant_slices} confirms the collocation of the large magnitude events.
For $Q_1$, a good correlation between the $P_{q_1} \left( \xi, \chi\right)$ and the magnitude is observed, which suggests a largely local flow response, and more importantly, $Q_1$ events are isolated at the upstream region of the coral roughness.
Comparing the $Q_2$ events further confirms the similar trend observed for $Q_1$, where the $P_{q_2} \left( \xi, \chi\right)$ are roughly collocated well with the regions where the peak value of the $Q_2$ events has a larger magnitude.
For all the staggered cases, the $Q_2$ events show a similar trend, upstream of the coral roughness, where the vortex tube stretching along the streamwise direction can be seen readily, where the $Q_2$ events dominate upstream of this vortex tube (see Fig. \ref{fig:quadrant_interactions}b).
While some differences can be observed for the three staggered cases with respect to the $Q_2$ event peaks, the overall spatial distribution of the peak magnitudes is similar.
A significant deviation from the overall trend is observed for the branching case that does not exhibit a strong upstream vortex, as a result, does not portray strong $Q_2$ events upstream of the coral roughness; also illustrated in Fig. \ref{fig:quadrant_interactions}c.
Flanked between the $Q_1$ and $Q_2$ contributions, the $Q_3$ events are primarily located at the front of the coral roughness, which can be attributed to turbulent fluxes in the wall direction.
The time-averaged flow along $\chi = 0.5$ between two consecutive coral roughnesses is shown in Fig. \ref{fig:vertical_streamlines}, which elucidates the overall trend observed for the spatial distribution of $P_{q_i} \left( \xi, \chi\right)$ as well as the magnitudes of the various quadrant contributions.
The time-averaged flow clearly illustrates the presence of the stagnation vortex \citep{Sadeh1980} that introduces largely local changes to the $Q_1$ and $Q_3$ events within the canopy, both in terms of small magnitude, large frequency events as well as large magnitude, tail events; by virtue of the upstream vortex tube as seen for all the cases in Fig. \ref{fig:quadrant_interactions}b-d.
Additionally, it is also clear to see that the mean flow between the staggered coral cases is observed to be relatively similar, while the stochastic case is vastly different due to the heterogeneity and spatial clusters.
This discussion illustrates the similarities observed for the various coral roughness types as detailed in the work by \cite{Hamilton2024} for all flow parameters of interest, and also underscores the nuanced differences observed between the branching and the massive cases, respectively.

\subsection{Stochastic Case}
\label{sec:stochastic_incanopy_stress}

\begin{figure}[hbpt!]
    \includegraphics[width=1\linewidth]{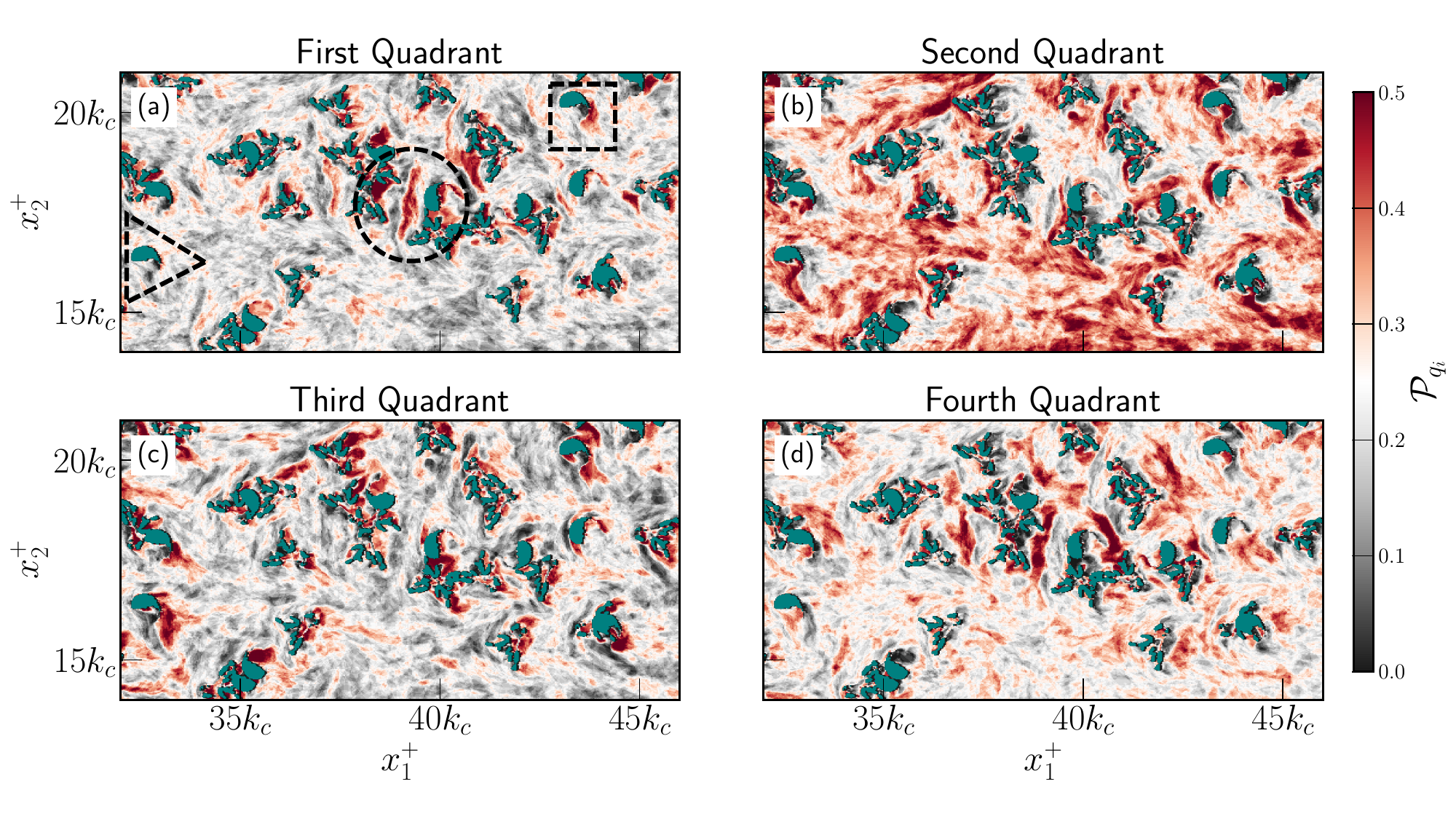}
    \caption{Comparison of the time-averaged quadrant probability ($P_{q_i} \left( \xi, \chi\right)$) for the stochastic cases within the canopy, close to the wall ($x_3^+ = 10$).}
    \label{fig:stochastic_quadrants}
\end{figure}

\begin{figure}[hbpt!]
    \includegraphics[width=1\linewidth]{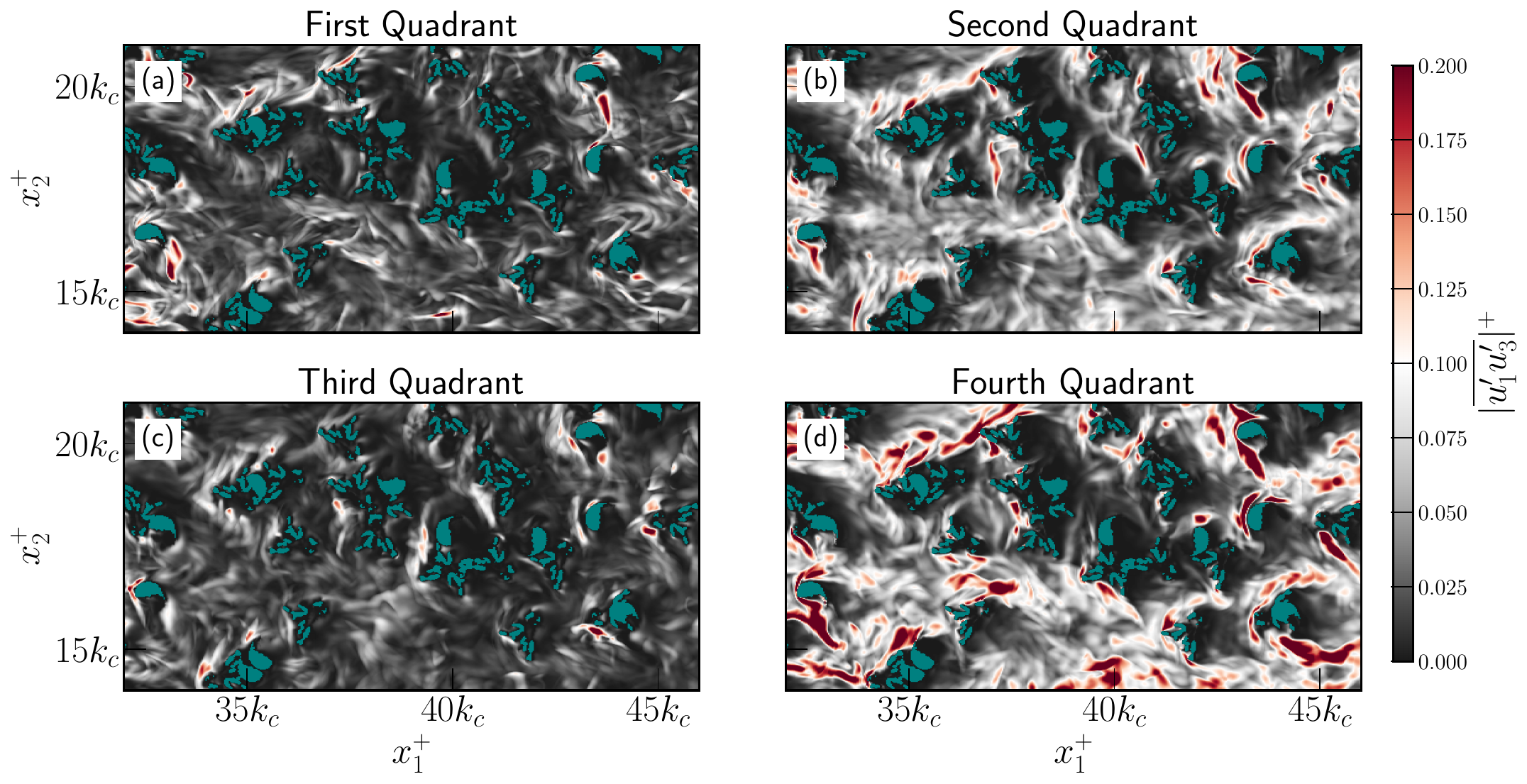}
    \caption{Comparison of the time-averaged quadrant magnitude for the stochastic cases within the canopy, close to the wall ($x_3^+ = 10$).}
    \label{fig:stochastic_mean_quadrants}
\end{figure}

As discussed in the previous sub-section, the staggered cases overall show similar behaviour when comparing the time-average flow and turbulent statistics, with some minor deviations by virtue of the underlying coral geometries.
However, introducing a stochastic spatial distribution using the same coral types (i.e., branching and massive) results in a vastly different time-averaged flow response as shown in Figs. \ref{fig:figure2} and \ref{fig:figure3}.
Unlike the staggered cases detailed in the previous sections, the stochastic case does not exhibit the spatial symmetries that can be leveraged to reduce the domain into pseudo-periodic tiles; thus, in this section, a small portion within the computational domain will be presented as shown in Figs. \ref{fig:stochastic_quadrants} and \ref{fig:stochastic_mean_quadrants}.
The stochastic case exhibits largely uncorrelated features when comparing $P_{q_i}(x_1^+,x_2^+)$ and the magnitude of the corresponding quadrants ($Q_i$) in space.
As marked by the dashed square in Fig. \ref{fig:stochastic_quadrants}a, while the probability $P_{q_1}(x_1^+,x_2^+)$ is relatively large for $Q_1$ events in the wake of the coral roughness, the corresponding $Q_1$ peak region as shown in Fig. \ref{fig:stochastic_mean_quadrants}a is offset in the south direction relative to the coral roughness.
As for the other quadrant events, a similar trend is observed where there is a lack of direct correlation between the large occurrence frequency and large magnitude of the respective quadrants as a function of space.
While this is not readily discussed as part of this work, the stochastic nature of the spatial arrangement introduces relatively more complex space-local interactions between the various quadrant events that distinguish this case from the staggered cases, as illustrated in this and the previous section.
Similar observations can be made for the regions marked with the dashed circle and the dashed triangle, suggesting that, irrespective of the occurrence frequency of the quadrant events, the peak magnitude for the corresponding $Q_i$ events is observed to showcase a vastly different behaviour compared to the staggered counterparts.
This key difference between the spatially regular and the stochastic spatial arrangement implies important consequences for the near-wall processes.
While scalar and sediment transport are not explicitly included in this work, using the turbulence dynamics as a proxy for these processes suggests largely local effects as a consequence of the spatial heterogeneity \citep{Mullarney2017,Blanco2001}.

\subsection{Plane-averaged quadrant contributions}

\begin{figure}[hbpt!]
    \includegraphics[width=1\linewidth]{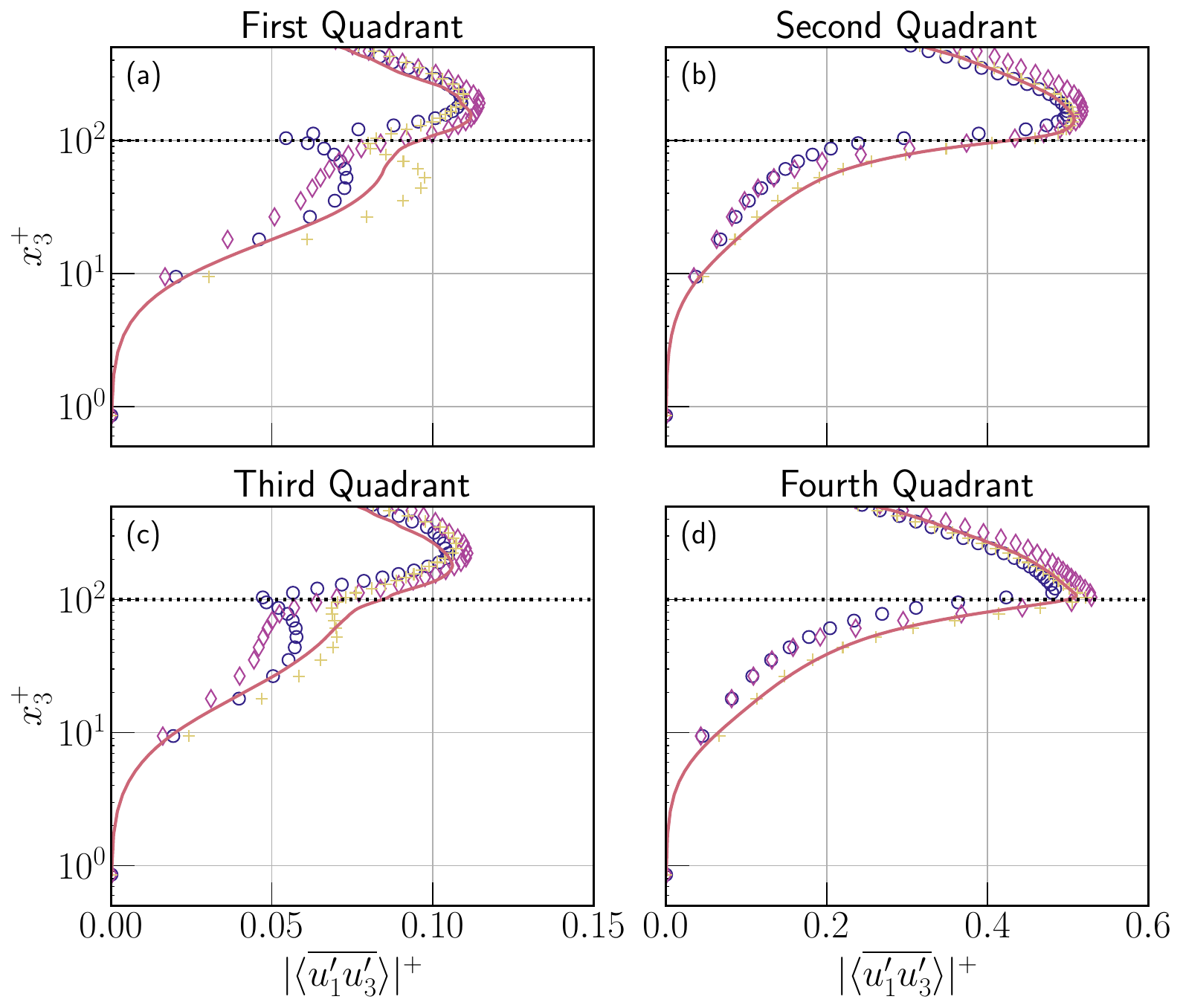}
    \caption{Comparison of the time- and plane-averaged quadrant contributions for the four cases considered in this work (absolute value). The horizontal black dotted line marks the crest of the coral roughness for all the cases. Data markers and colours identical to Fig. \ref{fig:figure2} and marked for one in ten data points.}
    \label{fig:quadrant_profiles}
\end{figure}

The spatial distribution of the various flow parameters of interest was discussed in the previous section; however, most often the context in which flow over coral reefs is studied requires relatively large-scale models and parameterisations \citep{Lowe2008,Rogers2018,Pomeroy2023}.
Consequently, the change as a function of the distance from the wall (i.e., vertical) becomes the primary quantity of interest as opposed to the spatial heterogeneity, which requires averaging over the spatial extent in the homogeneous flow directions.
To that end, despite the largely heterogeneous flow response described in the previous sections, Fig. \ref{fig:quadrant_profiles} compares the time- and plane-averaged quadrant contributions for the four cases mainly motivated by the intent to translate the findings to a more generalised case.
For the $Q_1$ events, there are some key differences observed within the coral canopy where the average response is similar for the cylinder and the massive, and the branching and the stochastic cases, respectively.
Specifically, the massive case exhibits the largest contribution within the canopy for $Q_1$ events, followed by the stochastic case, cylinder, and the branching case.
As the approximate shape of the massive case is similar to that of the cylinder (see Fig. \ref{fig:figure1}), the similarity observed between these two cases is not surprising.
As for the branching case, there is a relatively more gradual increase in the peak value of $Q_1$ within the canopy, but outside the canopy ($x_3^+ > 100$), the branching case is observed to overtake the other cases in terms of the peak magnitude just above the crest of the coral roughness.
Comparing the $Q_2$ magnitude as shown in Fig. \ref{fig:quadrant_profiles}b, the grouping observed for $Q_1$ is no longer identical to the stochastic and the massive cases are observed to have similar vertical variations and magnitudes within the coral canopy and above the canopy, while the branching and the cylinder cases show similar vertical profiles and magnitudes.
The overall trend for the $Q_3$ quadrant is similar to that of the $Q_1$ quadrant, while a similar clustering is also observed for the $Q_4$ quadrant in line with that of $Q_2$.
Further away from the wall (not shown in Fig. \ref{fig:quadrant_profiles}), there is good agreement between all the cases such that they have similar magnitudes suggesting that the turbulence statistics further away from the wall are more similar than within the coral canopy and in the vicinity of the coral roughness.

\begin{figure}[hbpt!]
    \includegraphics[width=1\linewidth]{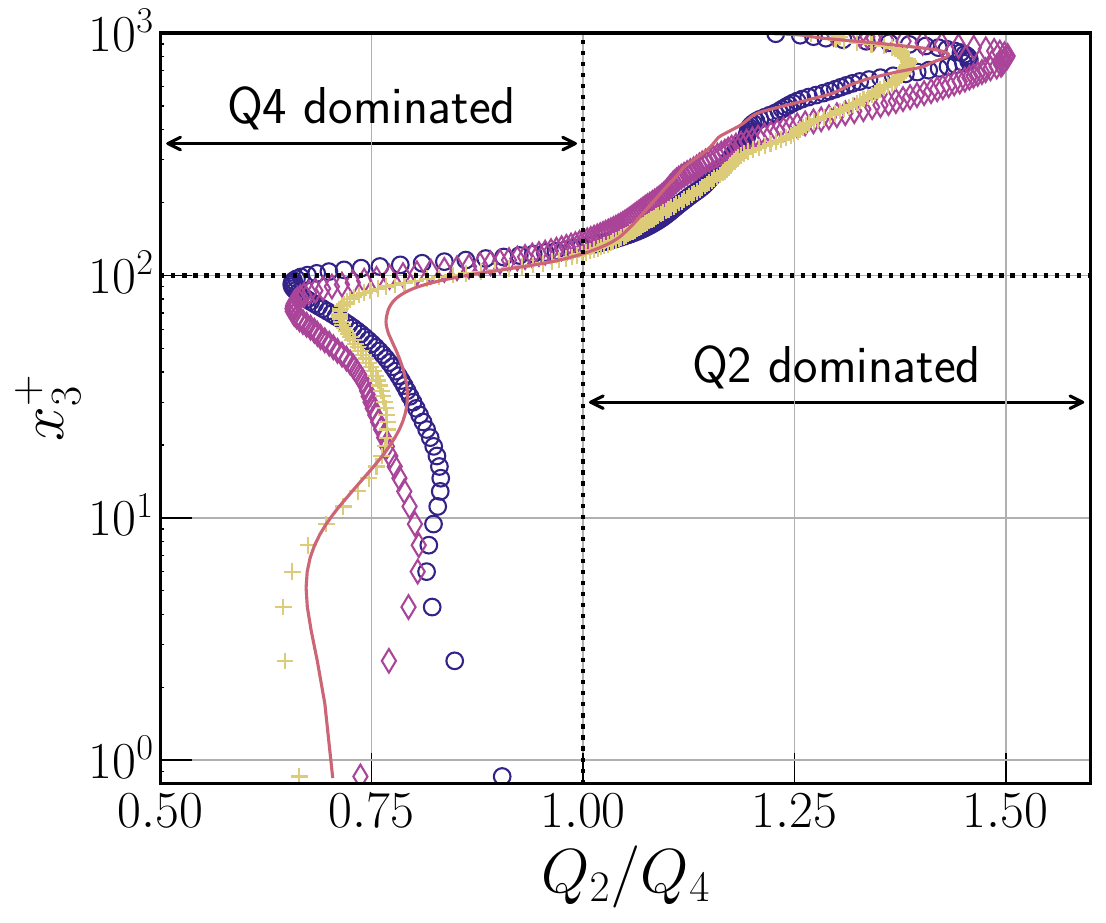}
    \caption{Comparison of the time- and plane-averaged ratio of the ejections and sweeps ($Q_2$ and $Q_4$) contributions as a function of the vertical coordinates. The vertical dotted line marks the magnitude where $Q_2$ and $Q_4$ have equal magnitude, while the horizontal dotted line marks the location of the crest of the coral roughness ($k_c$). The line and colour scheme is identical to Fig. \ref{fig:figure2} and the data markers are shown for every second data point.}
    \label{fig:q2q4_ratios}
\end{figure}

To further elucidate the relative importance of the primary contributors to the Reynolds stress (i.e., the $Q_2$ and $Q_4$ events), the ratio of ejections/sweep events is presented in Fig. \ref{fig:q2q4_ratios} for the four cases discussed in this work.
Within the canopy, sweep events are seen to dominate in line with the previous observations made by \cite{Katul1998,Finnigan2009} for atmospheric boundary layers over vegetated canopies.
Comparing the in-canopy magnitude across the four cases, the cylinder case shows a relatively balanced ratio of the sweep (flux towards the wall) and ejections (flux away from the wall), followed by the branching, stochastic, and massive cases, respectively.
For the massive case, this implies a relatively larger flux of higher momentum reaches closer to the wall within the canopy, closely followed by the stochastic case, as shown in Fig. \ref{fig:q2q4_ratios}.
The above detailed trend is observed for the very close to the bottom wall ($x_3^+ < 20$), and the trend reverses where the cylinder and the branching case exhibit a relatively larger flux toward the wall direction.
It is important to contextualise that, as shown in Fig. \ref{fig:quadrant_profiles}, the difference between the absolute magnitudes may be relatively small if not imperceptible. 
Above the coral canopy, the $Q_2$ quadrant contributions dominate the flow and largely reflect the trends observed for the Reynolds stress as compared in Fig. \ref{fig:figure3}.
These observations are in line with the previous findings made for atmospheric boundary layers over vegetated canopies \citep{Katul1998,Castro2006,Finnigan2009}, with some non-trivial effects seen for the stochastic case as detailed in this section.

\section{Consequences for Coastal Ocean Modelling}

\begin{figure}[hbpt!]
    \includegraphics[width=1\linewidth]{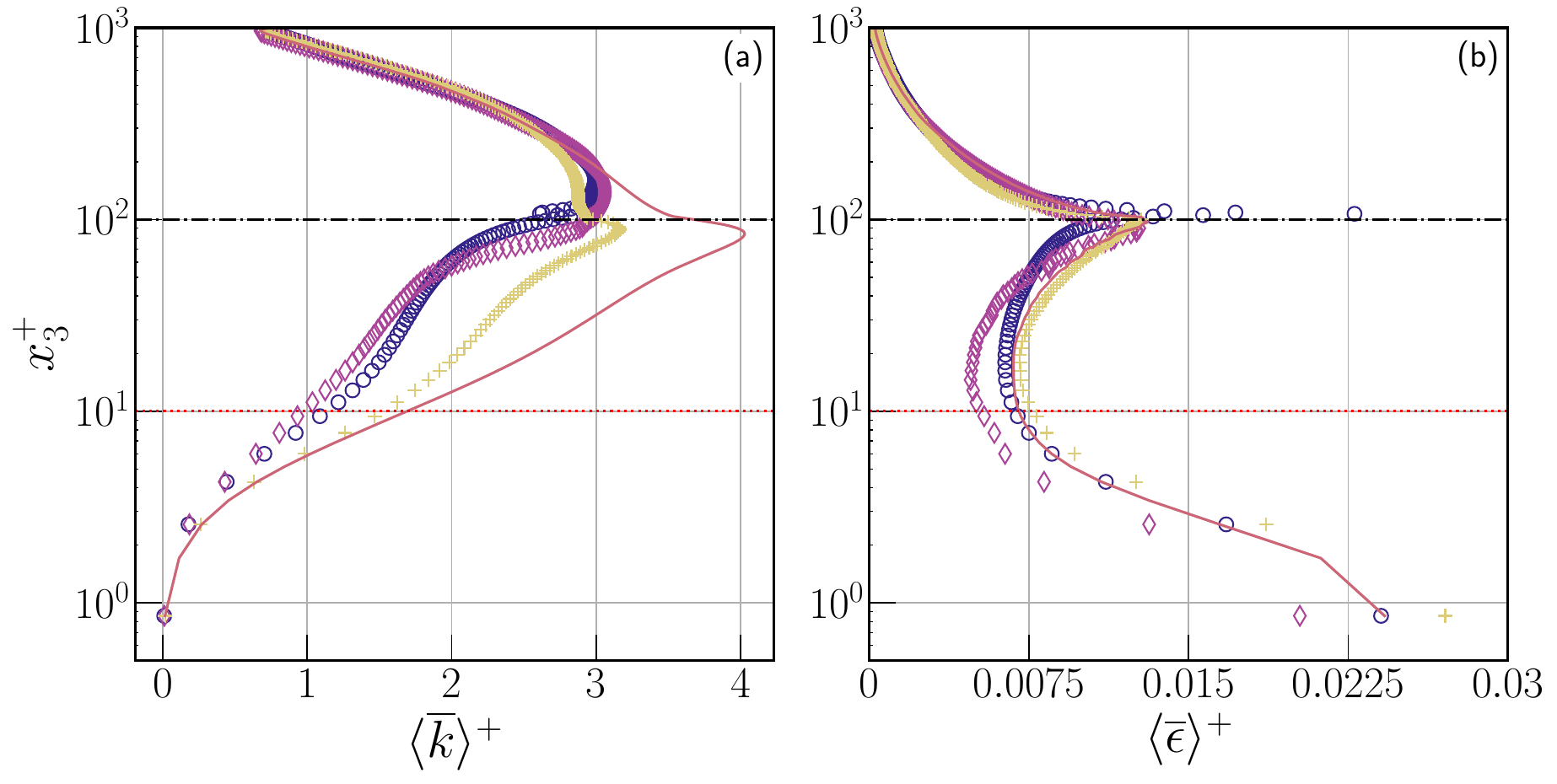}
    \caption{Comparison of the time- and plane-averaged vertical profiles for TKE (panel a) and TKE dissipation rate (panel b) for the four cases discussed in this work. The TKE profiles are non-dimensionalised using $u_{\tau}^2$ while the TKE dissipation rate is non-dimensionalised using $u_{\tau}^4/(\kappa \nu)$. The horizontal black dash-dot line corresponds to the crest of the coral roughness ($k_c$), while the red dotted line marks the location where the line and colour scheme are identical to Fig. \ref{fig:figure2} and the data markers are shown for every second data point.}
    \label{fig:tke_epsilon}
\end{figure}

In the previous sections, the discussion focused on quantifying the similarities and differences observed for the time-averaged flow and turbulent fluxes as a function of their spatial complexity (i.e., staggered vs. stochastic coral arrangements).
However, large-scale coastal ocean models \citep{Fringer2006,Zijlema2011} that are typically used to simulate coral canopies rely on parameterisations for the momentum and the turbulent kinetic energy (TKE) through the use of conventional two-equation type closures \citep{LaunderSharma1974,Mellor1982}.
Thus, despite the lack of conventional horizontal homogeneity for all the cases considered in this work, we present the TKE ($\langle \overline{k} \rangle$) and the TKE dissipation rate ($\langle \overline{\epsilon} \rangle$) as a function of vertical distance from the wall ($x_3^+$) in Fig. \ref{fig:tke_epsilon} to aid with modelling in-canopy closure models.
As compared in Fig. \ref{fig:tke_epsilon}a, the TKE profiles are observed to have the largest variation in magnitude within the coral canopy, with the stochastic case exhibiting the strongest TKE peak just under the coral roughness crest.
The overall trend for the massive case is observed to be similar to that of the stochastic case, except for the peak value of the TKE, which is 80\% of the peak magnitude observed for the stochastic case.
Between the four cases, the branching case is observed to exhibit the lowest levels of TKE throughout the canopy, closely followed by the cylinder case, where there is a close resemblance between the branching and the cylinder cases.
Above the coral canopy, the differences between the four cases rapidly vanish, and all the cases are observed to follow the same trend.
This is expected as the first-order effects are governed primarily by the mean roughness height ($k_c$) \citep{Jimenez2004}, which is identical for all four cases detailed in this work.
Consequently, further away from the coral canopy, the time-averaged flow responds identically when compared across the various coral canopies.
Overall, this suggests that despite the large variations observed for the time-averaged flow as detailed in Fig. \ref{fig:figure2} and the in-canopy turbulence levels (see Fig. \ref{fig:figure3}), the flow statistics further away from the wall are primarily governed by the mean roughness height.

When comparing the TKE dissipation rate (hereafter dissipation) vertical profiles across the four cases, most of the variations are also observed to be isolated within the coral canopy, while the above-canopy region for all the cases exhibits a relatively similar trend.
The cylinder case is observed to have a strong peak at the crest that is typically observed as there is a large shear layer for this coral roughness type (see Fig. \ref{fig:vorticity}).
A similar peak is observed for all the cases with a relatively lower magnitude when compared to the cylinder case at the coral roughness crest.
The overall differences between the various cases within the coral canopy are relatively small, suggesting that despite the largely heterogeneous response detailed in the previous sections, the flow features observed for all four cases are largely identical, with most of the differences isolated within the coral canopy.
Both the TKE and the dissipation profiles show identical behaviour above $x_3^+ \sim 300$, suggesting a universal response as a function of the coral roughness height ($k_c^+$).

One of the more important input parameters needed for modelling coral reefs is to quantify the effect of the underlying coral roughness on the flow through and above it. 
From the previous discussions, it is clear that the flow within the coral canopy is largely sensitive to the coral geometry.
More importantly, based on the spatial arrangement of the coral roughness, the flow within the coral canopy induces a largely local flow response that significantly varies for all the parameters of interest, such as the mean flow, TKE, dissipation, and turbulence levels.
Above the coral canopy, these differences are relatively less significant as shown in Fig. \ref{fig:tke_epsilon}, where the flow induces a relatively similar flow response within $x_3^+ \sim 300$ wall lengths above coral roughness geometry.
Consequently, it is clear that while stochastic coral roughness morphology introduces a large in-canopy response, the differences become relatively small with increasing distance from the wall for the small-scale turbulence processes.
This, however, does not entail that the impact of the coral morphology is negligible, as the small-scale, in-canopy differences observed can potentially have a large impact on the mean flow predictions as observed in Fig. \ref{fig:figure2}. 

\section{Conclusions}

In this work, we conducted a detailed comparative analysis of flow over four distinct rough-wall configurations, viz., staggered arrangement of cylinders (model for corals), branching type, and massive type coral roughness, and the stochastically arranged branching and massive coral roughnesses, in a unidirectional pressure-driven channel flow. 
Using a scale-resolving computational method, we resolved the requisite scales of motion and the coral roughness to gain insights into the similarities and differences observed between the four cases with varying geometric and spatial complexity.
Our data suggest that modelling the effect of heterogeneous coral canopies introduces a significant effect primarily within the coral canopy, similar to regular spatial arrangements of coral canopy flows \citep{Hamilton2024}, and a secondary effect on the flow above by inducing a relatively lower mean flow drag in the stochastic case, when compared to the staggered arrangement of coral roughnesses.
For the double-averaged flow statistics, most of the differences observed between the coral roughnesses are contained within the coral canopy, with the stochastic case exhibiting substantially different hydrodynamic response when compared with the staggered arrangements of coral roughnesses.
We also compared the spatial heterogeneity using the time-averaged flow and the contribution of the various Reynolds stress quadrants to illustrate the differences within the coral canopy for the various spatial configurations that lead to a strong local flow response that can become important for coastal boundary layer processes like sediment transport, and nutrient exchange within the boundary layer.

While this work sheds light on some of the dynamic distinctions between the various coral roughness configurations, our data and the requisite conclusions are limited by the associated computational cost and complexity metrics needed to fully generalise the findings to realistic coral canopies.
The numerous metrics that can be developed to quantify the coral geometric and spatial complexity can exceed tens of parameters \citep{brouwer2024}, thus limiting the use of a computationally expensive turbulence resolving framework as used in this work to adequately span the entire non-dimensional parameter space.
Additionally, for the staggered arrangements, this work used a fixed spacing ratio to emulate one of the many possible scenarios when considering the spatial arrangement of the coral roughnesses.
It is well known that depending on the relative spacing between the coral roughnesses and the coral height, the flow can significantly change. However, the overall dynamics are adequately captured using the parameters detailed in this work \citep{Jimenez2004}.
Our work also simulated the benthic bottom boundary layer using a quasi-static approximation that the pressure gradient drives the mean flow (see Fig. \ref{fig:reef_profile}) and does not contain the effect of the combined effect of mean-flow and wave motion as was done in \cite{Hamilton2024}.
However, this does severely limit translating the findings of this work, as the interaction between mean current and wave motions aims to introduce additional mean flow drag.
One of the most severe limitations of this work is the relatively low value of the friction Reynolds number when compared to the in-situ observed values that are approximately an order of magnitude higher.
While this is true, since the simulated friction Reynolds number is substantially large and is in the hydraulically rough wall flow regime, a similar flow response is expected for increasing Reynolds number as the flow response becomes asymptotic for increasing roughness Reynolds numbers \citep{Moddy1994}.

In conclusion, despite the relatively limited scope of our computational experiments, our work illustrated a substantial impact of the coral geometries on the double-averaged and time-averaged flow statistics. Our work illustrated that depending on the coral roughness geometry, arranged in an identical fashion, can introduce significant in-canopy differences and mean flow response.
Additionally, our data also suggests consequential local effects for staggered arranged coral roughnesses as well as the stochastically arranged coral roughness, which are important for relevant for modelling coral ocean bottom boundary layer dynamics and processes such as sediment transport, nutrient exchanges, and the fate of pollutants within complex coral canopies. 
Our findings provide important insights into the local hydrodynamics effects and their connections with the coral roughness geometry that can be used for simulating coral canopies with relatively more realistic scenarios, and also suggest important dynamical differences observed between the simple and complex coral configurations.
Overall, our data suggests that we should care about the geometric and spatial heterogeneity in coral reefs as they can introduce significant local effects as well as have important consequences for the mean flow statistics, as illustrated in our work.

\begin{acknowledgments}
AP would like to thank Nadine Hobeika for her insightful inputs and discussion during the preparation of this manuscript. AP would also like to thank Jenny Hamilton for the fruitful discussions and for her PhD thesis work, which largely inspired the work presented here. 
\end{acknowledgments}

\subsection*{Funding}
This work made use of the Dutch national e-infrastructure with the support of the SURF Cooperative, supplemented by the Dutch Research Council (NWO). This publication is part of the project “Multi-fidelity computational modelling of environmental fluid systems” with grant number 2024/ENW/01763969.

\subsection*{Carbon Footprint Statement}
This work made use of the Snellius supercomputer and had an estimated footprint of 1830~kg CO$_2$-equivalent (at least if not higher) using the Green Algorithms (\url{http://calculator.green-algorithms.org/}). This is equivalent to taking a 0.8 flight from New York (U.S.A) to Melbourne (Australia).

\subsection*{Declaration of generative AI and AI-assisted technologies in the writing process}
During the preparation of this work, the authors used Grammarly in order to proofread. After using this tool/service, the authors reviewed and edited the content as needed and would like to assume full responsibility for the content of the publication.

\subsection*{Data Availability Statement}

The simulations carried out in this work generated a total of 8TiB of data; thus, a section of the processed data has been made available to reproduce the results presented in this manuscript. This data can be accessed via the publicly accessible data repository (ADD 4TU DATA REPOSITORY BEFORE PUBLISHING)

%
%
\appendix

\section{Conditional Statistics: In-canopy Height Variations}
\label{sec:appendixA}

\begin{figure}[hbpt!]
    \includegraphics[width=1\linewidth]{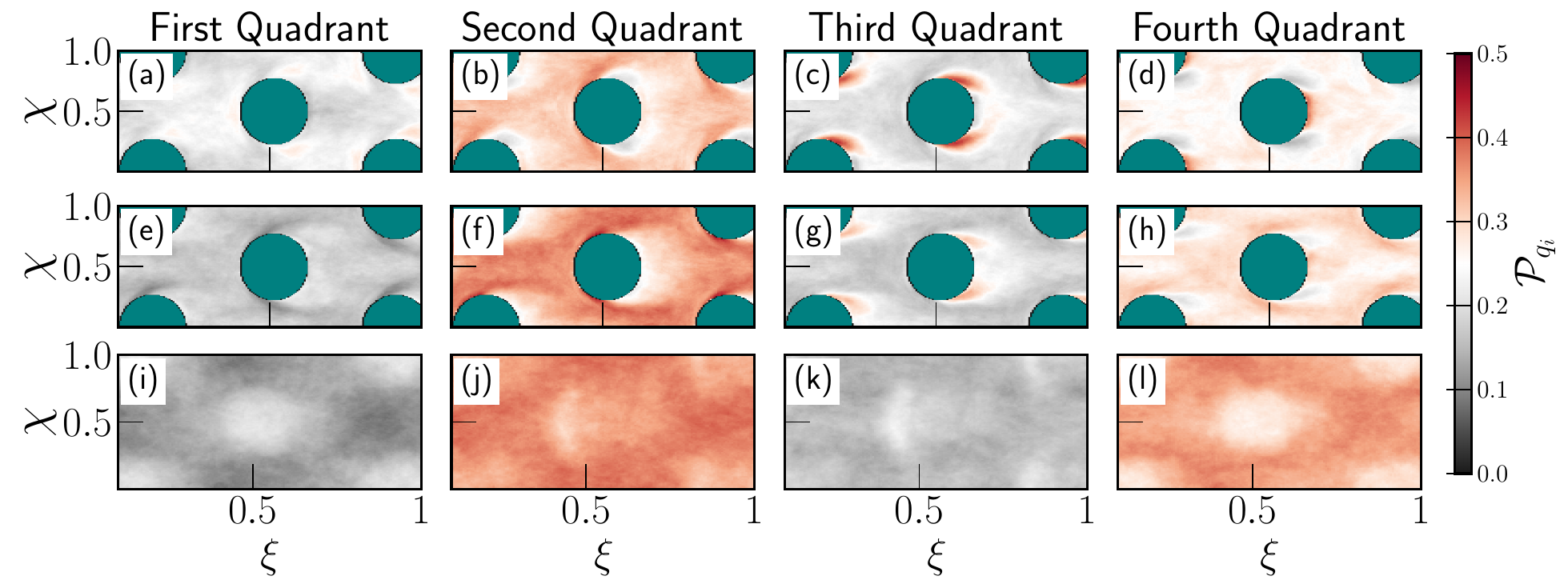}
    \caption{Time-averaged quadrant contributions for the cylinder case at three different heights above the wall. The first, second, and third rows correspond to $x_3^+ = 43$, $x_3^+ = 85$, and $x_3^+ = 120$, respectively. For all the panels shown here, each column corresponds to the quadrant contribution as mentioned in the title of the top row. The top two rows are contained within the canopy, while the last row is just outside the canopy layer, about 20 wall units above the coral roughness crest.}
    \label{fig:quadrant_cylinders}
\end{figure}

\begin{figure}[hbpt!]
    \includegraphics[width=1\linewidth]{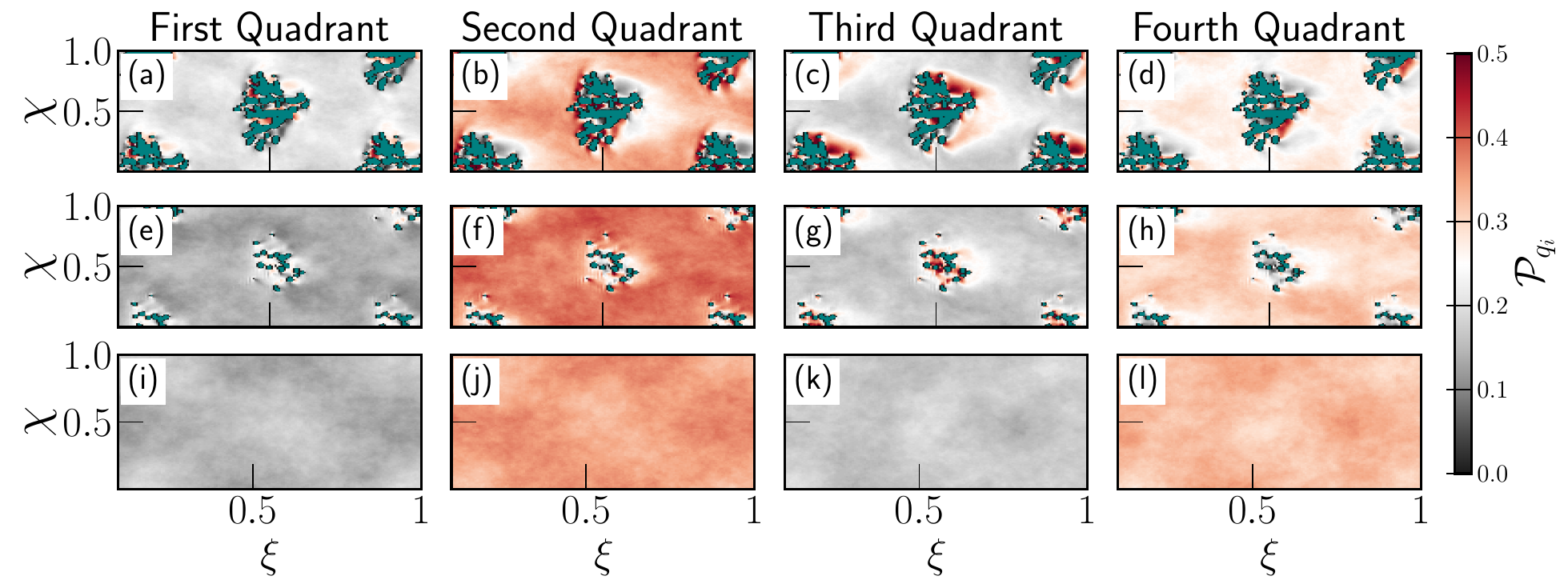}
    \caption{Time-averaged quadrant contributions for the branching case at three different heights above the wall. Data identical to Fig. \ref{fig:quadrant_cylinders}.}
    \label{fig:quadrant_branching}
\end{figure}

\begin{figure}[hbpt!]
    \includegraphics[width=1\linewidth]{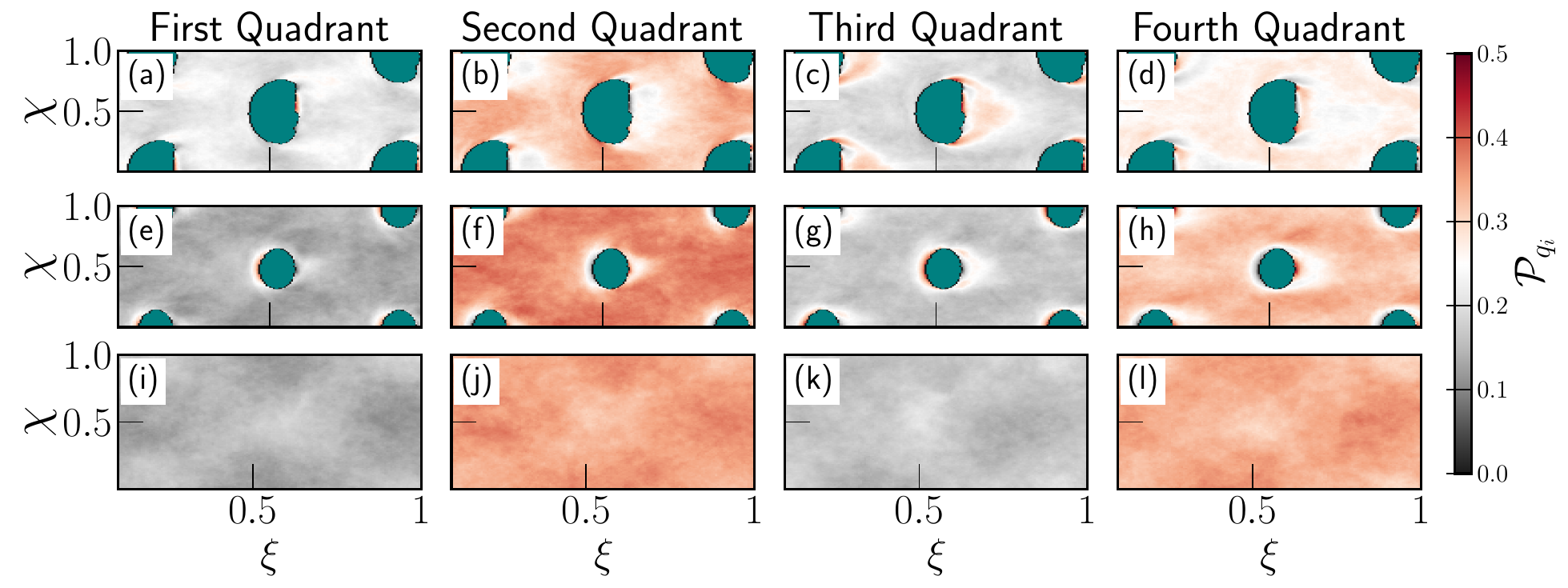}
    \caption{Time-averaged quadrant contributions for the massive case at three different heights above the wall. Data identical to Fig. \ref{fig:quadrant_cylinders}.}
    \label{fig:quadrant_massive}
\end{figure}

The contribution of the various quadrants to the Reynolds stress is also sensitive to the distance away from the wall \citep{NakagawaNezu1977,Pope2000,Wallace2016}. 
While the near-wall, in-canopy dynamics was discussed in Sec. \ref{sec:stres_within_canopy}, the sensitivity of the various quadrant contributions as a function of height is important to understand how the in-canopy dynamics is different from the outer layer.
Figs. \ref{fig:quadrant_cylinders}, \ref{fig:quadrant_branching}, and \ref{fig:quadrant_massive} compare the time-averaged quadrant contributions over the symmetric tiles (see Sec. \ref{sec:stres_within_canopy}) at three different heights above the wall (defined as $x_3 = 0.0$).
Based on the below presented height variations, there is clear evidence that with increasing distance away from the wall, the $Q_2$ and $Q_4$ events dominate in the frequency of occurrence when compared with the $Q_1$ and $Q_3$ events for all the staggered cases.
This is in line with previous findings of $ Q_2$ and $Q_4$ behaviour as a function of increasing distance away from the wall \citep{GerzSchumann1996,Pope2000,Wallace2016}.

Within the canopy, closer to the wall, there is a relatively larger abundance of $Q_1$ and $Q_3$ events primarily through their interaction with $Q_2$ and $Q_4$ events as detailed in \cite{GerzSchumann1996}.
At the leading front of the coral roughness (all cases), $Q_1$ events dominate in the vicinity of the vortex tube, which is observed to contribute substantially to these fluxes. 
As the distance from the wall increases, $Q_1$ events are observed less frequently, while small traces of $Q_3$ events (wall-ward interactions) can be observed for all cases detailed below.
Along the sides of the coral roughness (all cases), $Q_3$ events are observed at a relatively larger frequency as opposed to $Q_1$ events at the same height away from the wall, potentially hinting at interactions between the extended \textit{necklace} vortex \citep{Escauriaza2011} and the coral roughness that induce wall-ward interactions as shown in Fig. \ref{fig:quadrant_interactions}b.
This can be further confirmed by comparing the $Q_2$ and $Q_4$ events against the $Q_3$ events that flank the prior in spatial extent, while the latter is enclosed within.
It is also important to note that the $Q_1$ and $Q_3$ events' frequency of occurrence vanishes with increasing distance from the wall, while it has an observable magnitude within the canopy close to the coral roughness.
While these spatial trends only discuss the occurrence frequency and not the overall contribution towards the magnitude (e.g., less frequent yet extreme events) of the Reynolds stress, it is clear that the in-canopy response is substantially different from the above-canopy response.

\newpage

\bibliography{aipsamp}
\end{document}